\pdfoutput=1

\documentclass[11pt, table]{article}

\usepackage[preprint]{acl}

\usepackage{times}
\usepackage{latexsym}
\usepackage{subcaption}
\usepackage[T1]{fontenc}

\usepackage[utf8]{inputenc}

\usepackage{microtype}

\usepackage{inconsolata}

\usepackage{graphicx}

\usepackage{booktabs}
\usepackage{array}
\usepackage{highlight}
\usepackage{rotating}

\usepackage{graphicx}
\usepackage{tikz}
\usepackage{pgf}
\usepackage{array} 
\usepackage{tabularx} 
\usepackage{multirow}
\usepackage{adjustbox} 
\usetikzlibrary{shapes.geometric, arrows, positioning, fit, backgrounds}
\usepackage[most]{tcolorbox}
\usepackage{comment}  
\usepackage{multicol} 

\title{AIMSCheck: Leveraging LLMs for AI-Assisted Review of \\ Modern Slavery Statements Across Jurisdictions}

\author{
 \textbf{Adriana Eufrosina Bora\textsuperscript{1,2}\textsuperscript{*}},
 \textbf{Akshatha Arodi\textsuperscript{1}\textsuperscript{*}},
 \textbf{Duoyi Zhang\textsuperscript{2}},
 \textbf{Jordan Bannister\textsuperscript{1}},
\\
 \textbf{Mirko Bronzi\textsuperscript{1}},
 \textbf{Arsene Fansi Tchango\textsuperscript{1}},
 \textbf{Md Abul Bashar\textsuperscript{2}},
 \\
 \textbf{Richi Nayak\textsuperscript{2}},
  \textbf{Kerrie Mengersen\textsuperscript{2}}
\\
\centerline{$^1$Mila - Quebec AI Institute \quad $^2$The Queensland University of Technology} \\
\small
    \texttt{\{akshatha.arodi-nagaraja, jordan.bannister, m.bronzi, arsene.fansi.tchango\}@mila.quebec}\\
\small
    \texttt{\{a.bora, d25.zhang, m1.bashar, r.nayak, k.mengersen\}@qut.edu.au}
}

\begin{document}
\maketitle
\begingroup
\renewcommand\thefootnote{}\footnotetext{\textsuperscript{*}These authors contributed equally to this work.}
\endgroup
\begin{abstract}
Modern Slavery Acts mandate that corporations disclose their efforts to combat modern slavery, aiming to enhance transparency and strengthen practices for its eradication. However, verifying these statements remains challenging due to their complex, diversified language and the sheer number of statements that must be reviewed. The development of NLP tools to assist in this task is also difficult due to a scarcity of annotated data. Furthermore, as modern slavery transparency legislation has been introduced in several countries, the generalizability of such tools across legal jurisdictions must be studied. To address these challenges, we work with domain experts to make two key contributions. First, we present AIMS.uk and AIMS.ca, newly annotated datasets from the UK and Canada to enable cross-jurisdictional evaluation. Second, we introduce AIMSCheck, an end-to-end framework for compliance validation. AIMSCheck decomposes the compliance assessment task into three levels, enhancing interpretability and practical applicability. Our experiments show that models trained on an Australian dataset generalize well across UK and Canadian jurisdictions, demonstrating the potential for broader application in compliance monitoring. We release the benchmark datasets and AIMSCheck to the public to advance AI-adoption in compliance assessment and drive further research in this field.

\end{abstract}

\section{Introduction}

  \begin{figure}[ht]
     \centering
     \includegraphics[width=\linewidth]{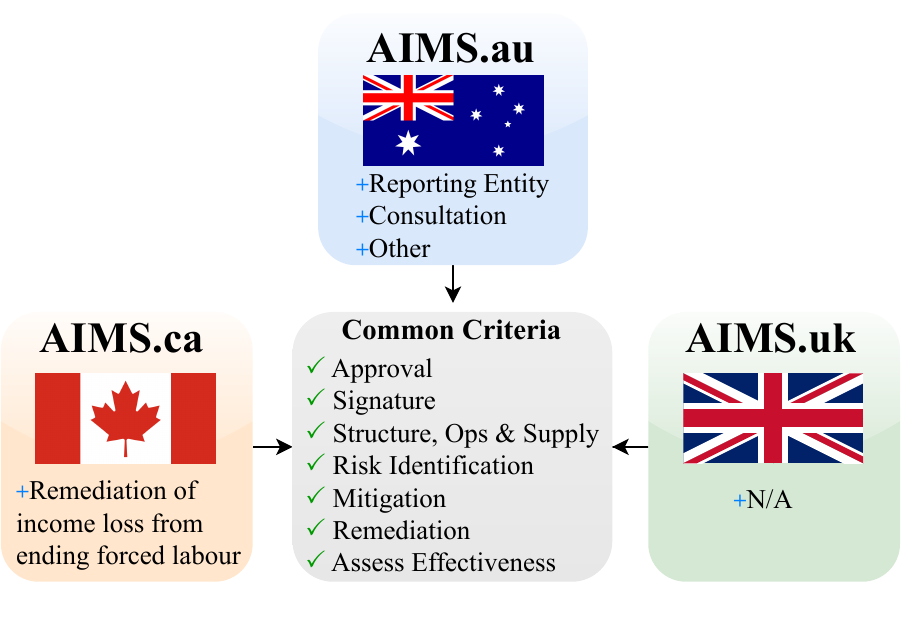}
     \caption{The common and unique mandatory reporting criteria associated with the Modern Slavery Acts in Australia, the United Kingdom, and Canada. The three datasets (AIMS.au, AIMS.uk, AIMS.ca) allow for cross-jurisdictional evaluation of compliance checking tools.}
     \label{fig:dataset-intro}
 \end{figure}

\begin{figure*}[h!]
    \centering
    \includegraphics[width=1\textwidth]{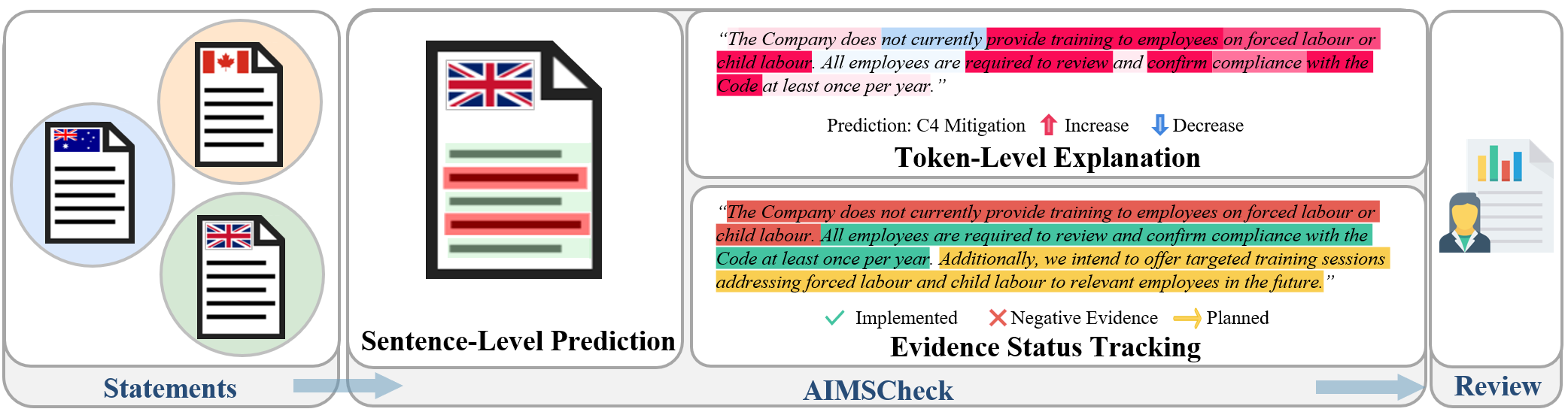}
    \caption{The AIMSCheck pipeline is designed to process modern slavery statements mandated by the Modern Slavery Acts of Australia, the United Kingdom and Canada. It generates predictions at sentence-level across multiple reporting criteria, complemented by token-level explainability techniques and evidence tracking. These outputs enable human analysts to efficiently and thoroughly review company compliance.}
    \label{fig:AIMSCheck.au}
\end{figure*}

Modern slavery affects over 50 million people worldwide, manifesting through exploitative practices such as forced labour, child labour or human trafficking, with 28 million forced into labour supplying global corporations \cite{free2022global}. To combat this, governments have enacted Modern Slavery Acts (MSAs), requiring large corporations to publish annual statements outlining their efforts to identify and mitigate modern slavery in their operations and supply chains. However, without robust monitoring and enforcement, forced labour remains undetected and unaddressed, allowing goods and services linked to exploitation to persist in global markets \cite{ChambersVastardis2020}. The UK (2015), Australian (2018), and Canadian (2024) MSAs mandate corporate disclosures, with approximately 12,000, 3,000, and 6,000 statements, respectively, expected to be annually published by corporations on their government registries \cite{ukgovmodernslaveryregistry, amsa2018_register, canadianmodernslaveryact}. These statements are detailed documents that should describe the companies, their efforts to address modern slavery, how they measure their success, and more. Compliance requirements vary by MSAs, defining specific criteria. An example statement with annotated criteria is in the Appendix 
 \ref{sec:appendix_example_statement}'s Figure~\ref{fig:appendix_example_statement}. Alongside historical data from \citet{bhrrcmodernslaveryregistry}, an estimated 80,000 statements exist globally, yet publicly available NLP-annotated datasets are scarce. Most research reviews only 100–200 statements manually\cite{WF2023, pham2023modern, dinshaw2022broken}, while larger efforts, like WikiRate’s 3,500 annotated statements \cite{WikiRate2023}, took over eight years to compile. This growing volume and slow review process make large-scale enforcement impractical.

This presents a compelling challenge for the NLP community to develop systems that support human reviewers in analysing and validating compliance statements. While existing methods focus on other legal tasks~\cite{santosh2024chronoslex,chalkidis-etal-2020-legal}, a gap remains in compliance checking for modern slavery statements. Unlike other legal classification tasks, this requires systems to precisely identify and extract relevant evidence while filtering out distractions, such as corporate jargon or vague assertions that lack substantive actions or pertinent information. Simple rule-based methods fail to capture these nuances~\cite{bora2019augmented}. The unstructured nature of mandated disclosures, without enforced document templates, further complicates this task.

Recently, Bora et al. \cite{bora2025} introduced AIMS.au, a dataset with sentence-level multi-label annotations aligned with the Australian MSA. While valuable, it raises several questions. It is unknown whether NLP models trained on Australian data will generalize across legal jurisdictions. Although MSAs share similarities, mapping compliance requires expertise, and manual annotation is resource-intensive. Domain experts unfamiliar with AI often hesitate to trust models, emphasizing the need for interpretability. Furthermore, compliance verification extends beyond relevance classification to evidence tracking, which involves monitoring future commitments and explicit denials related to compliance criteria. While relevance classification extracts any description relevant to a criterion as evidence, evidence tracking allows for a more nuanced analysis by distinguishing, for instance, between companies that already have a whistleblowing policy, those that do not, and those planning to implement one in the future. Without this capability, companies with varying levels of commitment may be treated equally, obscuring important insights. This distinction is essential for longitudinal studies, enabling the measurement of progress over time. With evidence tracking, reviewers can clearly identify which companies are improving and leading in compliance, rather than relying on static policy snapshots. This step is crucial for human validators to conduct a comprehensive assessment within a unified framework. The absence of an AI-assisted end-to-end compliance analysis system continues to hinder adoption.

\paragraph{Contributions} In this work, we address these gaps with two key contributions. First, we curate a jurisdictional mapping to evaluate generalizability across jurisdictions, as shown in Figure \ref{fig:dataset-intro}. To this end, we introduce the AIMS.uk and AIMS.ca datasets, derived from modern slavery statements collected from UK and Canadian government registries. These datasets consist of 50 statements from each jurisdiction that have been manually annotated by a domain expert. These diverse, well-structured datasets enable cross-jurisdictional evaluation of modern slavery disclosures. Our second contribution is the introduction of AIMSCheck (AI against Modern Slavery compliance Checks), an end-to-end framework designed to support human analysts in assessing compliance. As illustrated in Figure \ref{fig:AIMSCheck.au}, AIMSCheck operates at three distinct levels:
1) \textit{Sentence-Level}: classifies each sentence based on its relevance to compliance criteria 2) \textit{Token-Level}: enhances model transparency through explainability metrics, and 3) \textit{Evidence Status}: tracks sentences that support or refute the implementation, or future commitments of implementations, of measures, which we call evidence.

\begin{figure*}[htbp]
    \centering
     \includegraphics[width=\linewidth,trim={0 0px 0 0px},clip]{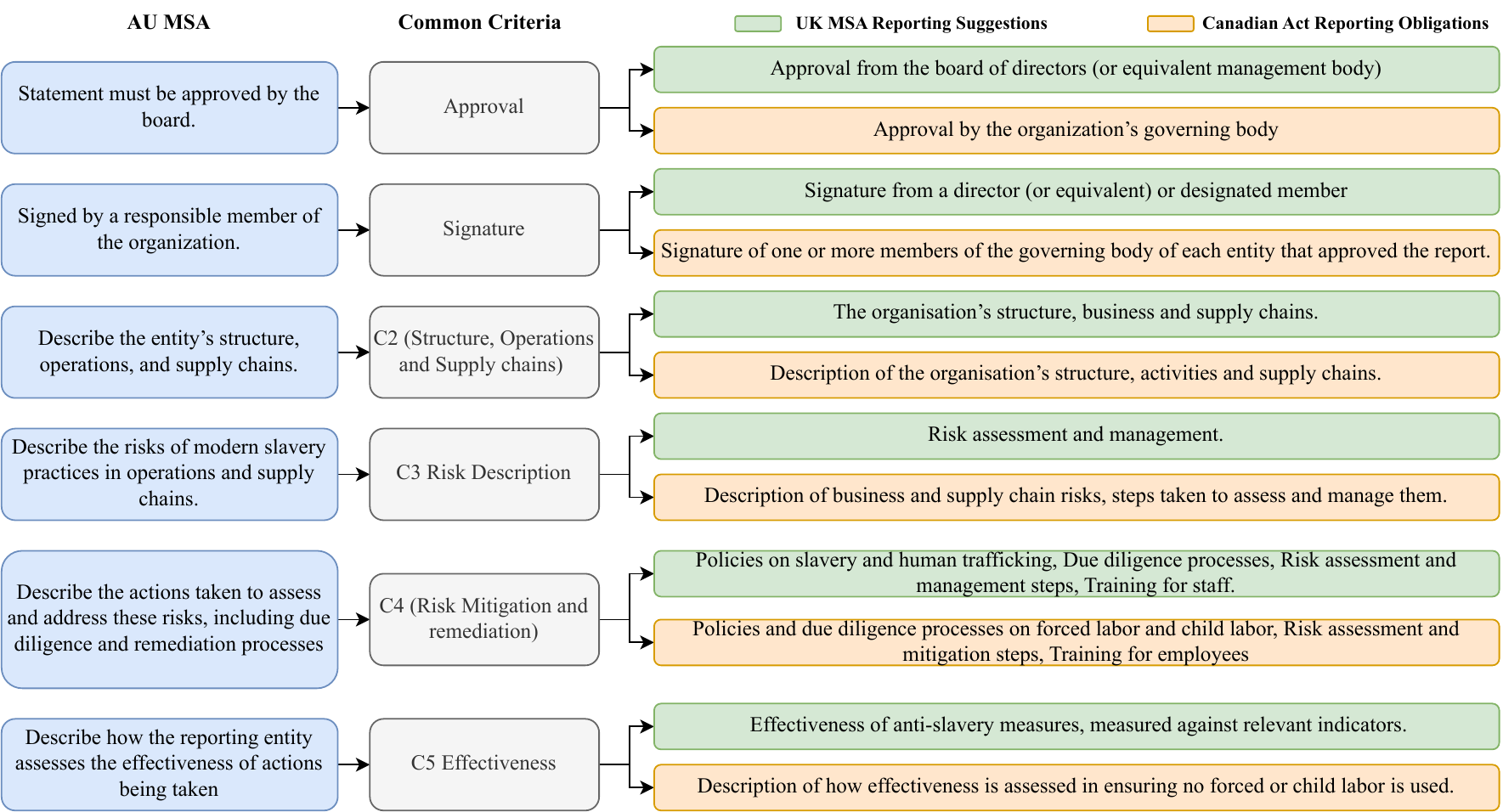}
    \caption{Mapping the AU MSA mandatory criteria, UK MSA reporting suggestions, and Canadian Act reporting obligations based on their common criteria.}
    \label{fig:data_mapping}
\end{figure*}

We experiment with zero-shot, few-shot, and fine-tuned models within the AIMSCheck framework. For sentence-level predictions, we fine-tune models on AIMS.au and evaluate them on AIMS.uk and AIMS.ca, finding strong cross-jurisdictional generalization. Fine-tuned models consistently outperform zero-shot and few-shot models across all datasets. However, all models struggle to distinguish relevant evidence for closely related compliance criteria. Developed in consultation with government and civil society organizations, this work aims to support supervisory agencies, such as Australia’s Attorney-General’s Department, the UK’s Home Office, and Public Safety Canada, in monitoring compliance. We also seek to help civil society and businesses develop evidence-based strategies to combat modern slavery. Finally, we urge the NLP community to advance methodologies for detecting modern slavery practices, addressing this urgent social issue.

\section{Datasets: AIMS.ca and AIMS.uk}

We introduce two new evaluation datasets in this work: AIMS.ca and AIMS.uk\footnote{\href{https://huggingface.co/datasets/mila-ai4h/AIMS.au}{Link to Hugging Face dataset.}}
. Each dataset contains 50 statements published by companies in Canada (CA) and the United Kingdom (UK), respectively, matching the size of the test set of AIMS.au, which includes 50 Australian statements. Each statement contains multiple pages. This results in a total of  2807 and 3658 sentences for AIMS.uk and AIMS.ca, respectively. To ensure comparability, we follow the same preprocessing and annotation guidelines as AIMS.au~\cite{bora2025}. The statements were sourced from online registries maintained by the Canadian government \cite{canadianmodernslaveryact} and the UK government \cite{ukgovmodernslaveryregistry}. Statements exceeding 200 sentences were removed, and from the remaining pool, 50 statements were randomly selected from each source for annotation. The selected texts were then split into sentences and annotated by an expert. To ensure high-quality annotations, a domain expert iteratively refined the labels until achieving a satisfactory level of confidence. A mapping between MSAs was curated by the domain experts by extracting nine common criteria across all MSAs as shown in Figure~\ref{fig:data_mapping}. A similar approach can be extended to broader human right legislations.

The datasets include detailed metadata for each statement, such as industry, company size, and publication year, enabling researchers to track corporate compliance trends, industry-specific reporting patterns, and regulatory impacts.

The AIMS.uk and AIMS.ca datasets include evidence-tracking annotations for two binary classification tasks: future action identification and negative evidence identification. Future action identification task identifies sentences where companies commit to future actions, where as negative evidence identification task capture sentences where companies acknowledge a lack of evidence or inaction. Notably, positive labels for both tasks are rare with 5.47\% in AIMS.uk and 1.12\% in AIMS.ca for future identification task, and 0.66\% and 2.76\% for negative evidence task in AIMS.uk and AIMS.ca respectively.

\begin{figure}[ht]
    \centering
    \includegraphics[width=\linewidth,trim={0 5px 0 5px},clip]{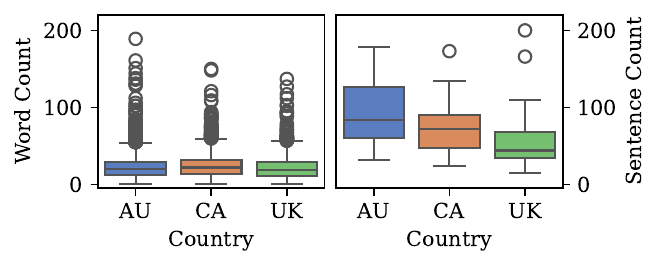}
    \caption{The distributions of word counts per sentence, and sentence counts per statement for the three datasets used in this work.}
    \label{fig:data_statistics}
\end{figure}
 
\begin{figure}[ht]
     \centering
     \includegraphics[width=\linewidth,trim={0 5px 0 5px},clip]{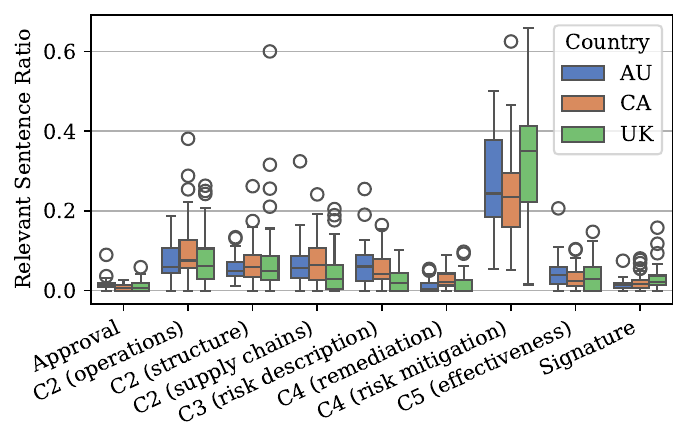}
     \caption{The distributions of the ratios of relevant sentences to total sentences, per statement, for the nine criteria and three datasets used in this work.}
     \label{fig:relevance_ratios}
 \end{figure}

We compare dataset statistics to assess their similarity and comparability. Figure \ref{fig:data_statistics} presents the distribution of sentence counts per statement and word counts per sentence across the annotated datasets. We observe that AIMS.uk and AIMS.ca mostly align with AIMS.au in word count per sentence, but they contain fewer sentences per statement. Figure \ref{fig:relevance_ratios} illustrates the ratio of relevant sentences to total sentences per statement in each dataset. Not all criteria are equally represented, criteria such as approval and signature have lower values since they are shorter and appear only once per statement, whereas C4 Mitigation has higher values due to its more descriptive nature. To ensure sufficient representation across criteria, Table \ref{tab:no_relevant_sentences_stats} reports the fraction of statements in each dataset containing at least one sentence addressing a given reporting criterion. While some criteria, such as Risk Mitigation, show similar distributions across datasets, others vary significantly. For the remediation criterion, the higher compliance rate (82\%) in Canada stems from a unique legal requirement mandating companies to address remediation of income loss caused by their actions, which broadens their discussion of remediation, unlike in Australia and the UK, where no such requirement exists, resulting in lower compliance (around 50\%).

\paragraph{Legislative Correspondence}
To facilitate cross-jurisdictional analysis, we curated a mapping between legislations building on existing work \cite{walkfree2022garment} and in consultation with diverse experts, identifying commonalities and differences among reporting criteria in Australia, Canada, and the UK as described below. Of the various criteria across these jurisdictions, the nine common ones were selected for this study,  as shown in Figure \ref{fig:dataset-intro} and further detailed in Figure \ref{fig:data_mapping}.

All legislations require statements to be approved and signed. They must describe the structure and supply chains of the reporting entity, outline identified risks and incidents, detail actions taken to mitigate risks and remediate incidents, and explain the processes for assessing the effectiveness of these actions. Some minor differences can be seen, for example "structure" and "supply chains" are consistent across all three regions, but they vary in how they refer to "operation" (Australia), "business" (UK), and "activities" (Canada). Moreover, the Australian law is more descriptive regarding risk mitigation, asking for broader criteria. In contrast, the UK and Canadian laws request more specific information, such as descriptions of policies, risk management practices, and employee training.

While the reporting suggestions of the UK Act are more generic and generally align with the others, the Australian Act mandates three unique reporting criteria: clear identification of the reporting entity, consultation process with owned and controlled entities in developing the statements, and any other relevant information. The Canadian Act specifies measures to address income loss for vulnerable families that resulted from their anti-slavery actions. An example statement with criteria and evidence status annotations is shown in Figure~\ref{fig:appendix_example_statement}.

\paragraph{Annotation Validation} We conducted an additional annotation validation process to further assess the reliability of the annotation. Another expert, who was deeply involved in developing the annotation guidelines, independently annotated 140 samples of the data. To validate our existing annotations, we used the samples to measured agreement. The Cohen’s Kappa score between the two annotators was 0.776 and Jaccard similarity was 0.813, demonstrating substantial agreement.

Many criteria align across jurisdictions, but some require one-to-many mappings or are jurisdiction-specific. Nine of Australia's eleven criteria have counterparts in Canada or the UK, while unique criteria need additional annotation. To achieve 1-to-1 mapping, we merged granular criteria into broader categories, particularly for mitigation, where Australia had a single broad criterion. This domain-expert-guided approach ensured consistency, and future work could refine annotations under top-level labels. The unique criteria for one jurisdiction, which do not apply to other jurisdictions, were not included in the study. More details on the exact text of the laws can be found in the Appendix \ref{appendix_mapping}.

\paragraph{Mapping Diverse Jurisdictions} Current business and human rights legislation can be categorized into two main types~\cite{li2024domestic}:

\textit{1) Mandatory reporting laws}, such as the Modern Slavery Acts in the UK, Australia, and Canada.

\textit{2) Mandatory human rights due diligence (mHRDD) laws}, exemplified by France’s Duty of Vigilance Law,  Germany’s Corporate Due Diligence Act, and Norway’s Transparency Act.

In this study, we focus on mandatory reporting laws, which benefit from centralized registries that facilitate data collection and analysis. In contrast, mHRDD laws often lack such registries and vary significantly in their enforcement mechanisms. As outlined in Appendix \ref{sec:other_jurisdictions}, it is feasible to map the criteria used in this study to the reporting elements within mHRDD laws. While certain elements may require adaptation, our criteria are sufficiently robust to accommodate diverse legal contexts. We anticipate that our framework will generalize across jurisdictions, as suggested by its cross-jurisdictional performance, and we encourage future work to build on these foundations. Also, extending this research to mHRDD laws poses challenges, particularly in collecting and annotating multilingual  documents, another promising area for future exploration.

\begin{table}[htbp]
    \centering
    \begin{tabular}{l m{1cm} m{1cm} m{1cm} m{1cm}}
        \toprule
        Question & AU & CA & UK \\
        \hline 
        Approval              & \g{0.98} & \g{0.52} & \g{0.51}\\
        C2 (operations)       & \g{0.98} & \g{0.98} & \g{0.96}\\
        C2 (structure)        & \g{1.00} & \g{0.96} & \g{0.84}\\
        C2 (supply chains)    & \g{0.96} & \g{0.90} & \g{0.75}\\
        C3 (risk description) & \g{0.92} & \g{0.96} & \g{0.55}\\
        C4 (remediation)      & \g{0.50} & \g{0.82} & \g{0.45}\\
        C4 (risk mitigation)  & \g{1.00} & \g{1.00} & \g{1.00}\\
        C5 (effectiveness)    & \g{0.86} & \g{0.84} & \g{0.69}\\
        Signature             & \g{0.90} & \g{0.80} & \g{0.82}\\
        \bottomrule
    \end{tabular}
    \caption{The fraction of statements within each dataset that include at least one sentence addressing the specified reporting criterion.}
    \label{tab:no_relevant_sentences_stats}
\end{table}

\section{AIMSCheck}
\label{sec:classification_experiments}

While a modern slavery compliance check might seem like a straightforward classification task, where each statement is simply labeled as compliant or not, such an approach is insufficient for real-world deployment. Civic society organizations and domain experts require a fair, transparent, and carefully monitored evaluation process \cite{Islam2022}. A simple fine-tuned model that outputs a binary decision lacks explainability, accountability, and the nuance needed for meaningful compliance assessment. Moreover, a simple pattern matching is insufficient for complex criteria like risk mititation, which can be descriptive and require background knowledge. 

To address these challenges, we propose AIMSCheck, an end-to-end framework that decomposes the compliance assessment task into multiple levels, ensuring a more interpretable and practical solution\footnote{\href{https://github.com/mila-ai4h/ai4h_aims-au}{Link to GitHub Repository.}}
. Instead of relying solely on automated classification, we incorporate a human review step, where experts make the final compliance decision based on structured assistance from our system as shown in Figure \ref{fig:AIMSCheck.au}. To enhance their efficiency and reliability, AIMSCheck introduces three key components as described below. By breaking down the compliance check into these structured steps, AIMSCheck improves adoption potential and provides decision support for human reviewers. A detailed example of the AIMSCheck workflow using a real statement is provided in Appendix \ref{appendix_realexample}.

\subsection{Sentence-Level Prediction}
We employ LLMs to classify each sentence as relevant or irrelevant to specific reporting criteria. We generate sentence-level predictions by considering the task as sentence-level multi-label binary classification problem which we evaluate across all nine criteria. We conduct our experiments using two open models—BERT \cite{devlin2018bert} and Llama3.2 (3B) \cite{llama3herd2024}—one closed model, OpenAI’s GPT-4o \cite{openai2023gpt4} and one open-weight model DeepSeek-R1 \cite{liu2024deepseek}. We evaluate GPT-4o in zero-shot, and chain-of-thought in zero-shot and few-shot settings, and compare its performance with BERT and Llama3.2 (3B), which are fine-tuned directly on the AIMS.au dataset. The fine tuned models were treated as binary classifiers for model evaluation and comparison, however, we note that these models may also be used as logistic regression models that output probabilities rather than class predictions. 

Our experiments follow two input data setups: (1) No context, where models classify the target sentence without any additional information, and (2) With context, where we include up to 100 words, balanced before and after the target sentence. This setup is based on prior work showing contextual benefits~\cite{bora2025}. To ensure robustness, we conducted experiments with different context lengths (0, 100, 200, 500) and found that 100 provided the best tradeoff between performance and computational efficiency. More details on experimental setup, including fine-tuning, training hyperparemeters zero-shot setup, prompts and examples are in Appendix \ref{sec:experimental_setting}.

\begin{figure*}
    \centering
      \includegraphics[width=\linewidth,clip]{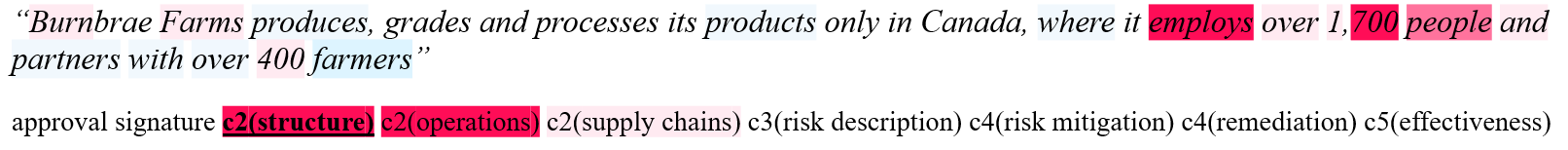}
        \caption{The SHAP plot illustrates the influence of tokens on the model's output. Here, we show the text (top) and the criteria labels (bottom). Red regions boost the model’s prediction, while blue regions suppress it, with colour intensity indicating influence strength. The model predicts a positive outcome for the C2 structure (underlined).}
        \label{fig:shap}
\end{figure*}

\subsection{Token-Level Explanation}
Sentence-level predictions are accompanied by token-level justifications to improve transparency. We use explainability methods \cite{linardatos2020explainable,lundberg2018explainable} to quantify the contribution of each token to the model's decision, to provide interpretable insights. Specifically, we use SHAP \cite{NIPS2017_7062}. SHAP is based on game theory and computes the impact of each token on model predictions by comparing model predictions with and without them. By attributing importance to specific tokens in corporate statements, SHAP values help human reviewers understand why each sentence is classified as compliant or non-compliant for a given category. This transparency fosters trust and aids in model refinement. 

\subsection{Evidence Status Tracking}
Evidence Status identifies cases where companies commit to future actions or explicitly state a lack of evidence or inaction, offering a fine-grained assessment of corporate responses to specific criteria. Given a sentence containing a detected criterion, our approach classifies it as indicating an already implemented action, a future action, or negative evidence. A future action signals intent to address the criterion, while negative evidence suggests inaction or denial. For future action detection, we use an NLTK-based tense classifier \cite{bird2006nltk}. For negative evidence detection, we apply a zero-shot BART-MNLI model \cite{lewis2019bart}, selecting the more probable hypothesis: whether a sentence downplays, denies, or avoids a criterion versus acknowledging it.

\begin{figure*}
        \centering
        \includegraphics[width=\linewidth]{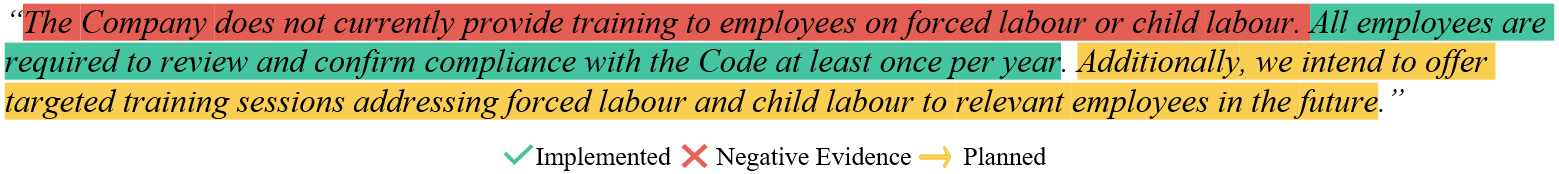}
        \caption{This example tracks the evidence status of a statement. The three underlined sentences are classified under the C4 mitigation criterion. 
        }
        \label{fig:future_negative}
\end{figure*}

\section{Results and Discussion}

\begin{table*}[ht]
    \centering
    \begin{tabular}{lcccc}
        \toprule
        Model & Context Words & AIMS.au & AIMS.ca & AIMS.uk \\
        \hline 
        Llama3.2 3B & 0   & 0.726 & 0.716 & 0.672 \\
        Llama3.2 3B & 100 & \textbf{0.738} & \textbf{0.719} & \textbf{0.686} \\
        BERT        & 0   & 0.694 & 0.677 & 0.653 \\
        BERT        & 100 & 0.719 & 0.700 & 0.669 \\
        \hline
        GPT-4o & 100  & 0.601 & 0.582 & 0.542 \\
        GPT-4o Chain-of-Thought & 100  & 0.559 & 0.560 & 0.500 \\
        GPT-4o Chain-of-Thought (few-shot) & 100  & 0.617 & 0.614 & 0.573  \\
        DeepSeek-R1 & 100  & 0.548 & 0.550 & 0.505  \\
        \bottomrule
    \end{tabular}
    \caption{Overall F1 evaluation results for fine-tuned (top) and zero/few-shot (bottom) models and across all three datasets on the sentence classification task. The overall F1 score is the average the over all the nine criteria. The best results in each column are in bold. DeepSeek-R1 is evaluated on AIMS.au.}
    \label{tab:macro_results}
\end{table*}

We evaluate sentence-level classification using the overall F1 score averaged across all nine criteria. Table \ref{tab:macro_results} shows results for all models across the three datasets. The results indicate that fine-tuned models consistently outperform zero-shot and few-shot models. Furthermore, a calibration analysis \cite{guo2017calibration} shows that the probabilities generated by the fine-tuned model accurately reflect the true likelihood of the model making a correct class prediction
(see Appendix \ref{sec:appendix_calibration}). Thus, human reviewers may also use probability values in their document review process as a way to assess the model's confidence in its predictions. We compare GPT-4o with DeepSeek-R1 and find that they achieve comparable performance\footnote{DeepSeek-R1 (2.51 bit quantized)}. With respect to the GPT-4o experiments, we observe a progression in performance from GPT-4o to Chain-of-Thought (CoT) prompting and further improvement when incorporating few-shot examples. This again highlights the complexity of the task and demonstrates that providing examples enhances performance. As expected, fine-tuned models trained on AIMS.au exhibit minor performance degradation when applied to UK and Canadian data. This suggests that while there is some shift between countries in compliance reporting, the gap is not substantial. 

\begin{table}[ht]
    \centering
    \begin{tabular}{l >{\centering\arraybackslash}m{.95cm} >{\centering\arraybackslash}m{.95cm} >{\centering\arraybackslash}m{.95cm}}
        \toprule
        Question & AU & CA & UK \\
        \hline 
        Approval              & \g{0.864} & \g{0.947} & \g{0.783}  \\
        C2 (operations)       & \g{0.769} & \g{0.803} & \g{0.789}  \\
        C2 (structure)        & \g{0.749} & \g{0.741} & \g{0.773}  \\
        C2 (supply chains)    & \g{0.805} & \g{0.656} & \g{0.704}  \\
        C3 (risk description) & \g{0.738} & \g{0.596} & \g{0.622}  \\
        C4 (remediation)      & \g{0.667} & \g{0.567} & \g{0.651}  \\
        C4 (risk mitigation)  & \g{0.669} & \g{0.674} & \g{0.646}  \\
        C5 (effectiveness)    & \g{0.592} & \g{0.526} & \g{0.525}  \\
        Signature             & \g{0.790} & \g{0.816} & \g{0.686}  \\
        \bottomrule
    \end{tabular}
    \caption{Per-criterion F1 evaluation results for the sentence classification task using Llama3.2 3B with 100 words of context.}
    \label{tab:detailed_results}

\end{table}
\begin{figure}[htpb]
    \centering
    \includegraphics[width=\linewidth,trim={0 25px 0 70px},clip]{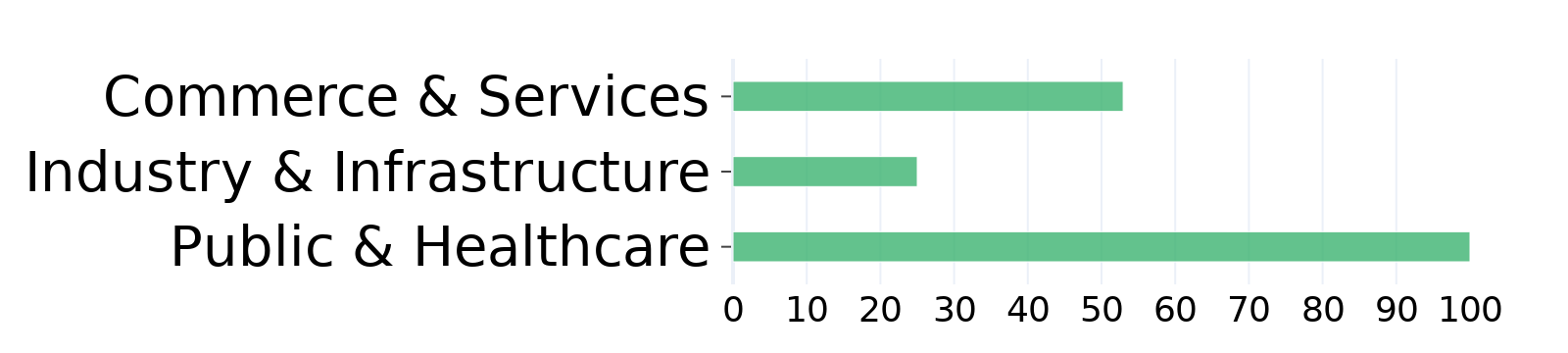}
    \caption{Proportion of Compliance for C3 (Risk Description) by Sector in AIMS.uk Dataset}
    \label{fig:sector}
\end{figure}

Table \ref{tab:detailed_results} reports F1 scores for the best-performing model (Llama3.2 3B with 100 words of context) across different reporting criteria. The results show significant variation across categories. The model performs well on relatively straightforward criteria, such as approval and signature, but struggles with more complex aspects like risk mitigation and remediation. This is expected, as criteria like approval are well-defined, as illustrated in Figure \ref{fig:data_mapping} and Figure \ref{fig:future_negative}, whereas risk mitigation is more subjective and can be vague.

For token-level explanation, we generate SHAP plots, which help identify the tokens that influenced the model’s predictions. Evaluating SHAP text and force plots for complex models is challenging \cite{carvalho2019machine, kokalj2021bert}. Therefore, we randomly sample a few examples and qualitatively analyse the plots. Figure \ref{fig:shap} shows a sentence annotated with SHAP values that allows users to see which tokens had a positive and negative influence on the classification decision. SHAP plots seem to capture the keywords influencing the prediction in many cases (see Appendix~\ref{sec:shap}).

For evidence status tracking, our model achieves macro-F1 scores of 66.83\% and 57.83\% for future prediction on AIMS.uk and AIMS.ca, respectively, while for the negative evidence task, the scores are 60.25\% and 67.80\%. While future evidence is relatively easy to predict, negative evidence is more challenging due to the convoluted language companies may use. Figure \ref{fig:future_negative} shows an example where the model correctly identifies both negative evidence (e.g., a company not providing training) and future promises.

\paragraph{Detailed Analysis} To better understand the characteristics of the AIMSCheck system, we conducted an in-depth error analysis of the best-performing models from fine-tuned and zero/few-shot category: Llama3.2-100 and GPT-4o-CoT Few-Shot, for the AIMS.uk and AIMS.ca datasets. It revealed that both models struggle with differentiating closely related criteria—such as distinguishing between description of a business' structure and operation—leading to significant error patterns. 
For example, elements from Criterion~2 (Structure, Operations, and Supply Chains) are frequently misclassified among themselves. Consider the following case:

\begin{quote}
\textit{“Jadestone is an independent oil and gas company focused on mid-life production and near-term development assets in the Asia Pacific region, operating principally in Australia, Indonesia, and Malaysia.”}
\end{quote}

In this example, the sentence contains relevant information about operations, yet the Llama 3.2 model erroneously classifies it as structure.

The GPT model, while capturing more information overall, tends to produce numerous false positives by introducing vague or misclassified content even with detailed prompts, whereas the Llama model’s conservative extraction approach results in a higher false-negative rate by omitting relevant sentences. Both models also face challenges with multi-criterion sentences, formatting issues (e.g., improper sentence segmentation and list extraction), and the overlooking of negative relevant statements. Minor annotation inconsistencies (1–2\% per criterion) further contribute to these errors. Additionally, DeepSeek-R1 follows prompts too rigidly compared to GPT-4o models. Detailed breakdown of error patterns and examples are in Appendix \ref{error_analysis}.

\paragraph{GPT CoT vs CoT (few-shot)} While providing few-shot examples improved overall performance of GPT CoT, at the criterion level, no consistent pattern emerged to suggest one method consistently outperforms the others as shown in Appendix \ref{sec:appendix_prompts}.

\paragraph{Modern Slavery Compliance Trends} We examined compliance trends in the AIMS.uk dataset using predictions from Llama 3.2 3B with a 100-context window, incorporating metadata. As an example of the insights users can gain from these predictions, Figure~\ref{fig:sector} illustrates that firms in the industry and infrastructure sectors tend to be less transparent in their risk disclosures. This relative opacity may suggest that these companies face lower external pressure to disclose risks compared to more public-facing industries like commerce and healthcare.
Appendix~\ref{sec:complianceTrend} provides a more detailed per-criteria analysis, showcasing the potential value of our work to assess statements at scale.

\begin{figure}
    \centering
    \includegraphics[width=\linewidth,trim={0 45px 0 105px},clip]{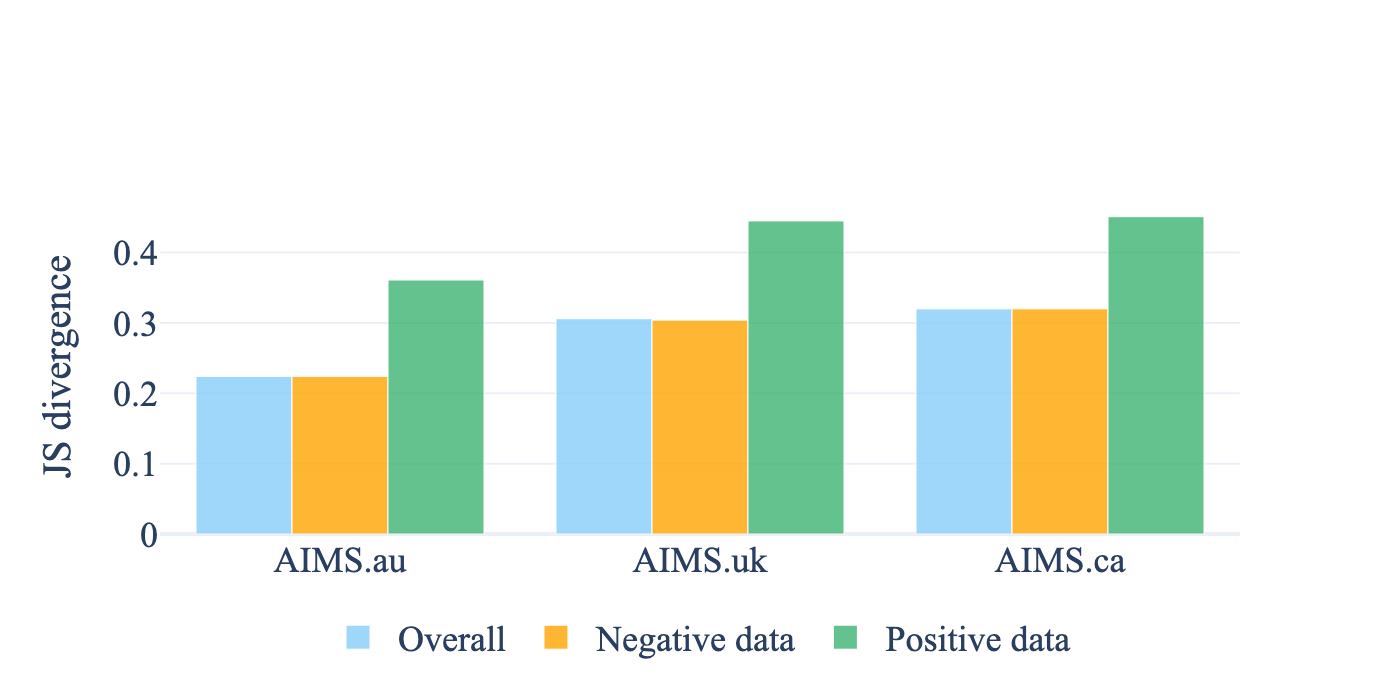}
    \caption{Jensen-Shannon (JS) divergences between AIMS.au (training) and testing vocabularies of all other datasets: Analysing overall, negative, and positive label-conditioned distributions across nine labels. Lower values indicate greater similarity.}
    \label{fig:JS_plot}
\end{figure}

\paragraph{Vocabulary Similarity} To understand why model performance remains consistent across datasets, we analyse Jensen-Shannon (JS) divergence \cite{dagan1997similarity, nielsen2019jensen}, which quantifies vocabulary overlap between probability distributions. A lower JS divergence (0–1 range) indicates greater similarity. Figure~\ref{fig:JS_plot} shows JS divergence between training (AIMS.au) and all other test datasets, with AIMS.au (test) serving as a baseline. The AIMS.uk and AIMS.ca datasets exhibit slightly higher divergence, suggesting minor cross-domain shifts. Notably, divergence is higher for positive samples, likely due to named entities such as company names and suppliers. Overall, vocabulary shifts between training and test sets remain minimal, explaining the stable model performance.

\section{Related Work}
While various models and datasets exist for related tasks such as legal or climate change claims classification \cite{webersinke2021climatebert, guha2023legalbench, chalkidis-etal-2020-legal}, few studies have applied machine learning to modern slavery statements, which present unique challenges. Prior work has used unsupervised topic modelling to analyse trends and terminology \cite{nersessian2022human, bora2019augmented}, but these methods were tested only on UK datasets and lack validation across different legislations. Moreover, they do not assess whether specific mandatory criteria are addressed, a key requirement for compliance monitoring. The release of AIMS.au \cite{bora2025} enables more precise analysis but remains limited to Australia. Our work differs by evaluating cross-jurisdictional generalization by introducing new datasets.

In recent years, NLP techniques have significantly advanced domain-specific text classification. Fine-tuning transformer-based language models, such as BERT~\cite{DBLP:conf/naacl/DevlinCLT19} and Llama~\cite{llama3herd2024}, and computationally inexpensive methods like LoRA (Low-Rank Adaptation)~\cite{DBLP:conf/iclr/XuXG0CZC0024} has proven effective for domain adaptation. Additionally, context-aware modelling has been shown to enhance performance in complex sentence-level tasks by providing complementary semantic information~\cite{tian-etal-2017-make, dara2023context, YANG2021157}. The rise of LLMs such as GPT-4~\cite{openai2023gpt4} and DeepSeek~\cite{liu2024deepseek} has popularized zero-shot and few-shot learning, which rely on minimal labeled data by leveraging LLMs’ reasoning abilities~\cite{NEURIPS2020_1457c0d6}. Effective prompt engineering plays a crucial role in guiding model inference, and Chain-of-Thought (CoT) prompting~\cite{wei2022chain} has further improved model reasoning by encouraging step-by-step logical processing~\cite{kojima2022large}. AIMSCheck integrates these state-of-the-art NLP techniques for modern slavery statement analysis. Furthermore, our work extends beyond sentence classification by introducing a comprehensive framework to support compliance tracking and monitoring.

\section{Conclusion}
In this work, we introduced two new datasets: AIMS.uk and AIMS.ca, enabling cross-jurisdictional generalization evaluation. Additionally, we introduced AIMSCheck, an end-to-end framework for modern slavery compliance checks. We evaluated its three components combining quantitative and qualitative assessments.  Our findings show that while GPT-4o models perform well, models fine-tuned on AIMS.au achieve better results across all jurisdictions on sentence-level classification. This suggests that with the right compliance mapping, domain adaptation for this task is possible. Further analysis highlights that certain compliance criteria, such as risk assessing effectiveness, are more difficult to classify, while others, like approval, are easier. Fine-tuned models tend to be conservative in predictions, whereas, Deepseek-R1 and GPT-4o models seem sensitive to prompts and few-shot examples. AIMSCheck represents a significant advancement in AI-assisted compliance assessment, streamlining the human review of modern slavery statements. Future work will focus on refining its components, improving model robustness, and expanding dataset coverage to support compliance monitoring across more jurisdictions. Ultimately, this research establishes the groundwork for AI-assisted, evidence-based policymaking and decision-making in the fight against modern slavery.

\section{Limitations}
We acknowledge several limitations of our work for analysing modern slavery statements.

Firstly, our evaluation is subject to dataset constraints and potential biases. The UK and Canadian datasets used for testing represent only small subsets of the available statements.
A comprehensive analysis of all available statements was beyond the scope of this study. Additionally, manual annotation was conducted by a single domain expert in modern slavery reporting. However, the expert adhered to the annotation specifications outlined by \cite{bora2025}, and iteratively refined the labels until achieving a satisfactory level of confidence.

Secondly, data formatting inconsistencies introduced errors that impacted performance as shown in the error analysis. The text preprocessing tools used in AIMSCheck, including Optical Character Recognition (OCR) and data parsers, occasionally failed on non-standard and complex documents, leading to incomplete sentences and lists and reducing the accuracy of the pipeline. Additionally, temporal drift \cite{santosh2024chronoslex} remains a concern, as legislation evolves over time, potentially affecting the relevance and accuracy of AIMSCheck’s outputs. Ongoing monitoring and adaptation will be necessary to maintain its effectiveness.

Thirdly, while this paper evaluates the effectiveness of AIMSCheck, its practical utility ultimately depends on how it is implemented in real-world applications. Also, SHAP plots are not always easy to interpret \cite{kokalj2021bert}. Future research should explore deployment strategies to ensure that AIMSCheck can be effectively integrated into compliance and transparency initiatives. A good starting place can be our indications of future directions in the error analysis (see Appendix \ref{error_analysis}).

Lastly, all datasets used in this research are exclusively in English, leaving the extension to multilingual scenarios as an avenue for future work.

\section{Ethical considerations}

In the development and evaluation of AIMSCheck, we have strictly relied on publicly available data submitted by organizations to governmental registries for public access and compliance checks.

We acknowledge that the annotation process, while guided by a domain expert with expertise across all three relevant legal frameworks, may introduce some degree of subjective bias. However, this risk is minimized through the expert’s extensive knowledge and multiple reiterations and quality checks for the annotations. 

Looking ahead, we note that deploying AIMSCheck at scale introduces risks such as misclassification, which could unintentionally affect the reputation of compliant organizations or result in missed opportunities to identify entities requiring further scrutiny—particularly if automated assessments are treated as definitive. To address this, we emphasize the importance of robust governance structures to oversee accountability, swiftly correct errors, and conduct detailed analyses of misclassifications. Given the complexity and contextual nuances of modern slavery statements, we also highlight the necessity of maintaining a human-in-the-loop approach to ensure careful interpretation of model outputs.

To foster transparency, trust, and accountability, we have implemented measures that enhance the interpretability of AIMSCheck’s decision-making process and clearly communicate its limitations. We advocate for ongoing development, responsible usage, and continuous refinement of this open-source resource. This should also include regular monitoring, periodic re-training, and adaptation to evolving legal and societal landscapes, ensuring the responsible and ethical deployment of AIMSCheck in practice.

\section*{Acknowledgments}

Part of this research was supported by the National Action Plan to Combat Modern Slavery 2020-25 Grants Program, administered by the Attorney-General’s Department. We thank Aldo Zaimi and Bruce Wen for their valuable discussions and feedback. We also extend our thanks to Jerome Solis for his support in managing this collaboration. 

\bibliography{custom}

\begin{thebibliography}{52}
\providecommand{\natexlab}[1]{#1}

\bibitem[{{Attorney-General's Department}(2023)}]{attorneygeneral2023}
{Attorney-General's Department}. 2023.
\newblock \href {https://modernslaveryregister.gov.au/resources/Commonwealth_Modern_Slavery_Act_Guidance_for_Reporting_Entities.pdf} {Commonwealth modern slavery act 2018: Guidance for reporting entities}.
\newblock Accessed: 2025-05-29.

\bibitem[{{Australian Government}(2025)}]{amsa2018_register}
{Australian Government}. 2025.
\newblock \href {https://modernslaveryregister.gov.au/} {{Modern Slavery Register}}.
\newblock Accessed on 09 February 2025.

\bibitem[{Bird(2006)}]{bird2006nltk}
Steven Bird. 2006.
\newblock Nltk: the natural language toolkit.
\newblock In \emph{Proceedings of the COLING/ACL 2006 Interactive Presentation Sessions}, pages 69--72.

\bibitem[{Bora et~al.(2025)Bora, St-Charles, Bronzi, Tchango, Rousseau, and Mengersen}]{bora2025}
Adriana~Eufrosiana Bora, Pierre-Luc St-Charles, Mirko Bronzi, Arsène~Fansi Tchango, Bruno Rousseau, and Kerrie Mengersen. 2025.
\newblock \href {https://arxiv.org/abs/2502.07022} {Aims.au: A dataset for the analysis of modern slavery countermeasures in corporate statements}.
\newblock \emph{Preprint}, arXiv:2502.07022.

\bibitem[{Bora(2019)}]{bora2019augmented}
Adriana-Eufrosina Bora. 2019.
\newblock \href {https://doi.org/10.13140/RG.2.2.15257.77921} {\emph{Using Augmented Intelligence in Accelerating the Eradication of Modern Slavery: Applied Machine Learning in Analyzing and Benchmarking the Modern Slavery Businesses Reports}}.
\newblock Dissertation.

\bibitem[{Brown et~al.(2020)Brown, Mann, Ryder, Subbiah, Kaplan, Dhariwal, Neelakantan, Shyam, Sastry, Askell, Agarwal, Herbert-Voss, Krueger, Henighan, Child, Ramesh, Ziegler, Wu, Winter, Hesse, Chen, Sigler, Litwin, Gray, Chess, Clark, Berner, McCandlish, Radford, Sutskever, and Amodei}]{NEURIPS2020_1457c0d6}
Tom Brown, Benjamin Mann, Nick Ryder, Melanie Subbiah, Jared~D Kaplan, Prafulla Dhariwal, Arvind Neelakantan, Pranav Shyam, Girish Sastry, Amanda Askell, Sandhini Agarwal, Ariel Herbert-Voss, Gretchen Krueger, Tom Henighan, Rewon Child, Aditya Ramesh, Daniel Ziegler, Jeffrey Wu, Clemens Winter, Chris Hesse, Mark Chen, Eric Sigler, Mateusz Litwin, Scott Gray, Benjamin Chess, Jack Clark, Christopher Berner, Sam McCandlish, Alec Radford, Ilya Sutskever, and Dario Amodei. 2020.
\newblock \href {https://proceedings.neurips.cc/paper_files/paper/2020/file/1457c0d6bfcb4967418bfb8ac142f64a-Paper.pdf} {Language models are few-shot learners}.
\newblock In \emph{Advances in Neural Information Processing Systems}, volume~33, pages 1877--1901. Curran Associates, Inc.

\bibitem[{{Business and Human Rights Resource Centre}(2025)}]{bhrrcmodernslaveryregistry}
{Business and Human Rights Resource Centre}. 2025.
\newblock Modern slavery registry.
\newblock \url{https://www.modernslaveryregistry.org/}.
\newblock Accessed on 09 February 2025.

\bibitem[{Carvalho et~al.(2019)Carvalho, Pereira, and Cardoso}]{carvalho2019machine}
Diogo~V Carvalho, Eduardo~M Pereira, and Jaime~S Cardoso. 2019.
\newblock Machine learning interpretability: A survey on methods and metrics.
\newblock \emph{Electronics}, 8(8):832.

\bibitem[{Chalkidis et~al.(2020)Chalkidis, Fergadiotis, Malakasiotis, Aletras, and Androutsopoulos}]{chalkidis-etal-2020-legal}
Ilias Chalkidis, Manos Fergadiotis, Prodromos Malakasiotis, Nikolaos Aletras, and Ion Androutsopoulos. 2020.
\newblock \href {https://doi.org/10.18653/v1/2020.findings-emnlp.261} {{LEGAL}-{BERT}: The muppets straight out of law school}.
\newblock In \emph{Findings of the Association for Computational Linguistics: EMNLP 2020}, pages 2898--2904, Online. Association for Computational Linguistics.

\bibitem[{Chambers and Vastardis(2020)}]{ChambersVastardis2020}
R.~Chambers and A.Y. Vastardis. 2020.
\newblock Human rights disclosure and due diligence laws: the role of regulatory oversight in ensuring corporate accountability.
\newblock \emph{Chicago Journal of International Law}, 21:323.

\bibitem[{Dagan et~al.(1997)Dagan, Lee, and Pereira}]{dagan1997similarity}
Ido Dagan, Lillian Lee, and Fernando Pereira. 1997.
\newblock Similarity-based methods for word sense disambiguation.
\newblock \emph{arXiv preprint cmp-lg/9708010}.

\bibitem[{Dara et~al.(2023)Dara, Srinivasulu, Babu, Ravuri, Paruchuri, Kilak, and Vidyarthi}]{dara2023context}
Suresh Dara, CH~Srinivasulu, CH~Madhu Babu, Ananda Ravuri, Tirumala Paruchuri, Abhishek~Singh Kilak, and Ankit Vidyarthi. 2023.
\newblock Context-aware auto-encoded graph neural model for dynamic question generation using nlp.
\newblock \emph{ACM transactions on asian and low-resource language information processing}.

\bibitem[{Deepseek-R1-GGUF()}]{unslothdeepseek}
Unsloth Deepseek-R1-GGUF.
\newblock \href {https://huggingface.co/unsloth/DeepSeek-R1-GGUF} {Unsloth deepseek-r1-gguf}.

\bibitem[{Devlin(2018)}]{devlin2018bert}
Jacob Devlin. 2018.
\newblock Bert: Pre-training of deep bidirectional transformers for language understanding.
\newblock \emph{arXiv preprint arXiv:1810.04805}.

\bibitem[{Devlin et~al.(2019)Devlin, Chang, Lee, and Toutanova}]{DBLP:conf/naacl/DevlinCLT19}
Jacob Devlin, Ming{-}Wei Chang, Kenton Lee, and Kristina Toutanova. 2019.
\newblock \href {https://doi.org/10.18653/V1/N19-1423} {{BERT:} pre-training of deep bidirectional transformers for language understanding}.
\newblock In \emph{Proceedings of the 2019 Conference of the North American Chapter of the Association for Computational Linguistics: Human Language Technologies, {NAACL-HLT} 2019, Minneapolis, MN, USA, June 2-7, 2019, Volume 1 (Long and Short Papers)}, pages 4171--4186. Association for Computational Linguistics.

\bibitem[{Dinshaw et~al.(2022)Dinshaw, Nolan, Sinclair, Marshall, McGaughey, Boersma, Bhakoo, Goss, and Keegan}]{dinshaw2022broken}
Freya Dinshaw, Justine Nolan, Amy Sinclair, Shelley Marshall, Fiona McGaughey, Martijn Boersma, Vikram Bhakoo, Jasper Goss, and Peter Keegan. 2022.
\newblock \href {https://www.hrlc.org.au/reports-news-commentary/broken-promises} {Broken promises: Two years of corporate reporting under australia’s modern slavery act}.
\newblock Technical Report.

\bibitem[{{Federal Republic of Germany}(2021)}]{germany2021}
{Federal Republic of Germany}. 2021.
\newblock \href {https://www.bmas.de/EN/Services/Publications/a712-gesetz-unternehmerische-sorgfaltspflichten-lieferketten.html} {Act on corporate due diligence obligations in supply chains (lieferkettensorgfaltspflichtengesetz)}.
\newblock Accessed: 2025-05-29.

\bibitem[{Guha et~al.(2023)Guha, Nyarko, Ho, Ré, Chilton, Narayana, Chohlas-Wood, Peters, Waldon, Rockmore, Zambrano, Talisman, Hoque, Surani, Fagan, Sarfaty, Dickinson, Porat, Hegland, Wu, Nudell, Niklaus, Nay, Choi, Tobia, Hagan, Ma, Livermore, Rasumov-Rahe, Holzenberger, Kolt, Henderson, Rehaag, Goel, Gao, Williams, Gandhi, Zur, Iyer, and Li}]{guha2023legalbench}
Neel Guha, Julian Nyarko, Daniel~E. Ho, Christopher Ré, Adam Chilton, Aditya Narayana, Alex Chohlas-Wood, Austin Peters, Brandon Waldon, Daniel~N. Rockmore, Diego Zambrano, Dmitry Talisman, Enam Hoque, Faiz Surani, Frank Fagan, Galit Sarfaty, Gregory~M. Dickinson, Haggai Porat, Jason Hegland, Jessica Wu, Joe Nudell, Joel Niklaus, John Nay, Jonathan~H. Choi, Kevin Tobia, Margaret Hagan, Megan Ma, Michael Livermore, Nikon Rasumov-Rahe, Nils Holzenberger, Noam Kolt, Peter Henderson, Sean Rehaag, Sharad Goel, Shang Gao, Spencer Williams, Sunny Gandhi, Tom Zur, Varun Iyer, and Zehua Li. 2023.
\newblock \href {https://arxiv.org/abs/2308.11462} {Legalbench: A collaboratively built benchmark for measuring legal reasoning in large language models}.
\newblock \emph{Preprint}, arXiv:2308.11462.

\bibitem[{Guo et~al.(2017)Guo, Pleiss, Sun, and Weinberger}]{guo2017calibration}
Chuan Guo, Geoff Pleiss, Yu~Sun, and Kilian~Q Weinberger. 2017.
\newblock On calibration of modern neural networks.
\newblock In \emph{International conference on machine learning}, pages 1321--1330. PMLR.

\bibitem[{{Home Office}(2024)}]{homeoffice2024}
{Home Office}. 2024.
\newblock \href {https://assets.publishing.service.gov.uk/media/67dd67b4c6194abe97358c26/Transparency_in_supply_chains_a_practical_guide.pdf} {Transparency in supply chains: A practical guide}.
\newblock Accessed: 2025-05-29.

\bibitem[{Islam and Staden(2022)}]{Islam2022}
Muhammad~Azizul Islam and Chris J.~Van Staden. 2022.
\newblock \href {https://doi.org/10.1007/s10551-021-04878-1} {Modern slavery disclosure regulation and global supply chains: Insights from stakeholder narratives on the uk modern slavery act}.
\newblock \emph{Journal of Business Ethics}, 180(2):455--479.

\bibitem[{Kojima et~al.(2022)Kojima, Gu, Reid, Matsuo, and Iwasawa}]{kojima2022large}
Takeshi Kojima, Shixiang~Shane Gu, Machel Reid, Yutaka Matsuo, and Yusuke Iwasawa. 2022.
\newblock Large language models are zero-shot reasoners.
\newblock \emph{Advances in neural information processing systems}, 35:22199--22213.

\bibitem[{Kokalj et~al.(2021)Kokalj, {\v{S}}krlj, Lavra{\v{c}}, Pollak, and Robnik-{\v{S}}ikonja}]{kokalj2021bert}
Enja Kokalj, Bla{\v{z}} {\v{S}}krlj, Nada Lavra{\v{c}}, Senja Pollak, and Marko Robnik-{\v{S}}ikonja. 2021.
\newblock Bert meets shapley: Extending shap explanations to transformer-based classifiers.
\newblock In \emph{Proceedings of the EACL hackashop on news media content analysis and automated report generation}, pages 16--21.

\bibitem[{Lewis(2019)}]{lewis2019bart}
Mike Lewis. 2019.
\newblock Bart: Denoising sequence-to-sequence pre-training for natural language generation, translation, and comprehension.
\newblock \emph{arXiv preprint arXiv:1910.13461}.

\bibitem[{Li and Xiang(2024)}]{li2024domestic}
Zhuolun Li and Yu~Xiang. 2024.
\newblock Domestic mandatory human rights due diligence laws as global business and human rights regulation.
\newblock \emph{Am. U. Int'l L. Rev.}, 40:319.

\bibitem[{Linardatos et~al.(2020)Linardatos, Papastefanopoulos, and Kotsiantis}]{linardatos2020explainable}
Pantelis Linardatos, Vasilis Papastefanopoulos, and Sotiris Kotsiantis. 2020.
\newblock Explainable ai: A review of machine learning interpretability methods.
\newblock \emph{Entropy}, 23(1):18.

\bibitem[{Liu et~al.(2024)Liu, Feng, Xue, Wang, Wu, Lu, Zhao, Deng, Zhang, Ruan et~al.}]{liu2024deepseek}
Aixin Liu, Bei Feng, Bing Xue, Bingxuan Wang, Bochao Wu, Chengda Lu, Chenggang Zhao, Chengqi Deng, Chenyu Zhang, Chong Ruan, et~al. 2024.
\newblock \href {https://arxiv.org/abs/2412.19437} {Deepseek-v3 technical report}.
\newblock \emph{arXiv preprint arXiv:2412.19437}.

\bibitem[{Llama~Team(2024)}]{llama3herd2024}
AI~@~Meta Llama~Team. 2024.
\newblock \href {https://arxiv.org/abs/2407.21783} {The llama 3 herd of models}.
\newblock \emph{arXiv preprint arXiv:2407.21783}.

\bibitem[{Lundberg and Lee(2017)}]{NIPS2017_7062}
Scott~M Lundberg and Su-In Lee. 2017.
\newblock \href {http://papers.nips.cc/paper/7062-a-unified-approach-to-interpreting-model-predictions.pdf} {A unified approach to interpreting model predictions}.
\newblock In I.~Guyon, U.~V. Luxburg, S.~Bengio, H.~Wallach, R.~Fergus, S.~Vishwanathan, and R.~Garnett, editors, \emph{Advances in Neural Information Processing Systems 30}, pages 4765--4774. Curran Associates, Inc.

\bibitem[{Lundberg et~al.(2018)Lundberg, Nair, Vavilala, Horibe, Eisses, Adams, Liston, Low, Newman, Kim et~al.}]{lundberg2018explainable}
Scott~M Lundberg, Bala Nair, Monica~S Vavilala, Mayumi Horibe, Michael~J Eisses, Trevor Adams, David~E Liston, Daniel King-Wai Low, Shu-Fang Newman, Jerry Kim, et~al. 2018.
\newblock Explainable machine-learning predictions for the prevention of hypoxaemia during surgery.
\newblock \emph{Nature Biomedical Engineering}, 2(10):749.

\bibitem[{Nersessian and Pachamanova(2022)}]{nersessian2022human}
David Nersessian and Dessislava Pachamanova. 2022.
\newblock Human trafficking in the global supply chain: Using machine learning to understand corporate disclosures under the uk modern slavery act.
\newblock \emph{Harv. Hum. Rts. J.}, 35:1.

\bibitem[{Nielsen(2019)}]{nielsen2019jensen}
Frank Nielsen. 2019.
\newblock On the jensen--shannon symmetrization of distances relying on abstract means.
\newblock \emph{Entropy}, 21(5):485.

\bibitem[{{Norwegian Consumer Authority}(2022)}]{forbrukertilsynet2022}
{Norwegian Consumer Authority}. 2022.
\newblock The transparency act.
\newblock \url{https://www.forbrukertilsynet.no/vi-jobber-med/apenhetsloven/the-transparency-act}.
\newblock Accessed: 2025-05-29.

\bibitem[{OpenAI()}]{openaigpt4o}
OpenAI.
\newblock \href {https://platform.openai.com/docs/models/gpt-4o} {Openai: Gpt4o}.

\bibitem[{OpenAI(2023)}]{openai2023gpt4}
OpenAI. 2023.
\newblock \href {https://arxiv.org/abs/2303.08774} {Gpt-4 technical report}.
\newblock \emph{arXiv preprint arXiv:2303.08774}.

\bibitem[{{Parliament of the United Kingdom}(2015)}]{ModernSlaveryAct2015}
{Parliament of the United Kingdom}. 2015.
\newblock Modern slavery act 2015.
\newblock \url{https://www.legislation.gov.uk/ukpga/2015/30/contents/enacted}.
\newblock Accessed: 2025-02-13.

\bibitem[{Pham et~al.(2023)Pham, Cui, and Ruthbah}]{pham2023modern}
Nga Pham, Bei Cui, and Ummul Ruthbah. 2023.
\newblock \href {https://www.monash.edu/business/mcfs/our-research/all-projects/modern-slavery/modern-slavery-statement-disclosure-quality} {Modern slavery disclosure quality: {ASX100} companies update {FY2022} modern slavery statements}.

\bibitem[{{Public Safety Canada}(2024)}]{publicsafety2024}
{Public Safety Canada}. 2024.
\newblock \href {https://www.publicsafety.gc.ca/cnt/cntrng-crm/frcd-lbr-cndn-spply-chns/prpr-rprt-en.aspx} {Fighting against forced labour and child labour in canadian supply chains: Preparing for the 2025 reporting cycle}.
\newblock Accessed: 2025-05-29.

\bibitem[{{Public Safety Canada}(2025)}]{canadianmodernslaveryact}
{Public Safety Canada}. 2025.
\newblock Fighting against forced labour and child labour in supply chains act.
\newblock \url{https://www.publicsafety.gc.ca/cnt/rsrcs/lbrr/ctlg/rslts-en.aspx?l=7}.
\newblock Accessed on 09 February 2025.

\bibitem[{{République Française}(2017)}]{france2017official}
{République Française}. 2017.
\newblock Loi n° 2017-399 du 27 mars 2017 relative au devoir de vigilance des sociétés mères et des entreprises donneuses d'ordre.
\newblock \url{https://www.legifrance.gouv.fr/jorf/id/JORFTEXT000034290626/}.
\newblock Journal Officiel de la République Française, Accessed: 2025-05-29.

\bibitem[{Santosh et~al.(2024)Santosh, Vuong, and Grabmair}]{santosh2024chronoslex}
TYS Santosh, Tuan-Quang Vuong, and Matthias Grabmair. 2024.
\newblock Chronoslex: Time-aware incremental training for temporal generalization of legal classification tasks.
\newblock \emph{arXiv preprint arXiv:2405.14211}.

\bibitem[{Tian et~al.(2017)Tian, Yan, Mou, Song, Feng, and Zhao}]{tian-etal-2017-make}
Zhiliang Tian, Rui Yan, Lili Mou, Yiping Song, Yansong Feng, and Dongyan Zhao. 2017.
\newblock \href {https://doi.org/10.18653/v1/P17-2036} {How to make context more useful? an empirical study on context-aware neural conversational models}.
\newblock In \emph{Proceedings of the 55th Annual Meeting of the Association for Computational Linguistics (Volume 2: Short Papers)}, pages 231--236, Vancouver, Canada. Association for Computational Linguistics.

\bibitem[{{UK Government}(2025)}]{ukgovmodernslaveryregistry}
{UK Government}. 2025.
\newblock Modern slavery statement registry.
\newblock \url{https://modern-slavery-statement-registry.service.gov.uk/}.
\newblock Accessed on 009 February 2025.

\bibitem[{{Walk Free}(2022{\natexlab{a}})}]{walkfree2022garment}
{Walk Free}. 2022{\natexlab{a}}.
\newblock Beyond compliance in the garment industry.
\newblock \url{https://tinyurl.com/y6yxrjwb}.
\newblock Accessed on 05 January 2025.

\bibitem[{{Walk Free}(2022{\natexlab{b}})}]{free2022global}
{Walk Free}. 2022{\natexlab{b}}.
\newblock \href {https://www.ilo.org/media/370826/download} {Global estimates of modern slavery: Forced labour and forced marriage}.
\newblock Technical Report, International Labour Organization (ILO).

\bibitem[{{Walk Free}(2023)}]{WF2023}
{Walk Free}. 2023.
\newblock \href {https://cdn.walkfree.org/content/uploads/2023/09/20162158/WF-Beyond-Compliance-Renewable-Energy-Sector.pdf} {Beyond compliance: Renewable energy sector}.

\bibitem[{Webersinke et~al.(2021)Webersinke, Kraus, Bingler, and Leippold}]{webersinke2021climatebert}
Nicolas Webersinke, Mathias Kraus, Julia~Anna Bingler, and Markus Leippold. 2021.
\newblock Climatebert: A pretrained language model for climate-related text.
\newblock \emph{arXiv preprint arXiv:2110.12010}.

\bibitem[{Wei et~al.(2022)Wei, Wang, Schuurmans, Bosma, Xia, Chi, Le, Zhou et~al.}]{wei2022chain}
Jason Wei, Xuezhi Wang, Dale Schuurmans, Maarten Bosma, Fei Xia, Ed~Chi, Quoc~V Le, Denny Zhou, et~al. 2022.
\newblock Chain-of-thought prompting elicits reasoning in large language models.
\newblock \emph{Advances in neural information processing systems}, 35:24824--24837.

\bibitem[{{WikiRate}(2025)}]{WikiRate2023}
{WikiRate}. 2025.
\newblock \href {https://wikirate.org/UK_Modern_Slavery_Act_Research} {{UK} modern slavery act research}.
\newblock Data Repository.

\bibitem[{Wolf et~al.(2020)Wolf, Debut, Sanh, Chaumond, Delangue, Moi, Cistac, Rault, Louf, Funtowicz, Davison, Shleifer, von Platen, Ma, Jernite, Plu, Xu, Scao, Gugger, Drame, Lhoest, and Rush}]{wolf2020huggingfacestransformersstateoftheartnatural}
Thomas Wolf, Lysandre Debut, Victor Sanh, Julien Chaumond, Clement Delangue, Anthony Moi, Pierric Cistac, Tim Rault, Rémi Louf, Morgan Funtowicz, Joe Davison, Sam Shleifer, Patrick von Platen, Clara Ma, Yacine Jernite, Julien Plu, Canwen Xu, Teven~Le Scao, Sylvain Gugger, Mariama Drame, Quentin Lhoest, and Alexander~M. Rush. 2020.
\newblock \href {https://arxiv.org/abs/1910.03771} {Huggingface's transformers: State-of-the-art natural language processing}.
\newblock \emph{Preprint}, arXiv:1910.03771.

\bibitem[{Xu et~al.(2024)Xu, Xie, Gu, Chen, Chang, Zhang, Chen, Zhang, and Tian}]{DBLP:conf/iclr/XuXG0CZC0024}
Yuhui Xu, Lingxi Xie, Xiaotao Gu, Xin Chen, Heng Chang, Hengheng Zhang, Zhengsu Chen, Xiaopeng Zhang, and Qi~Tian. 2024.
\newblock \href {https://openreview.net/forum?id=WvFoJccpo8} {Qa-lora: Quantization-aware low-rank adaptation of large language models}.
\newblock In \emph{The Twelfth International Conference on Learning Representations, {ICLR} 2024, Vienna, Austria, May 7-11, 2024}. OpenReview.net.

\bibitem[{Yang et~al.(2021)Yang, Wang, Wong, Shi, and Tu}]{YANG2021157}
Baosong Yang, Longyue Wang, Derek~F. Wong, Shuming Shi, and Zhaopeng Tu. 2021.
\newblock \href {https://doi.org/10.1016/j.neucom.2021.06.009} {Context-aware self-attention networks for natural language processing}.
\newblock \emph{Neurocomputing}, 458:157--169.

\end{thebibliography}

\appendix

  \begin{figure*}[!htb]
     \centering
     \includegraphics[width=\linewidth]{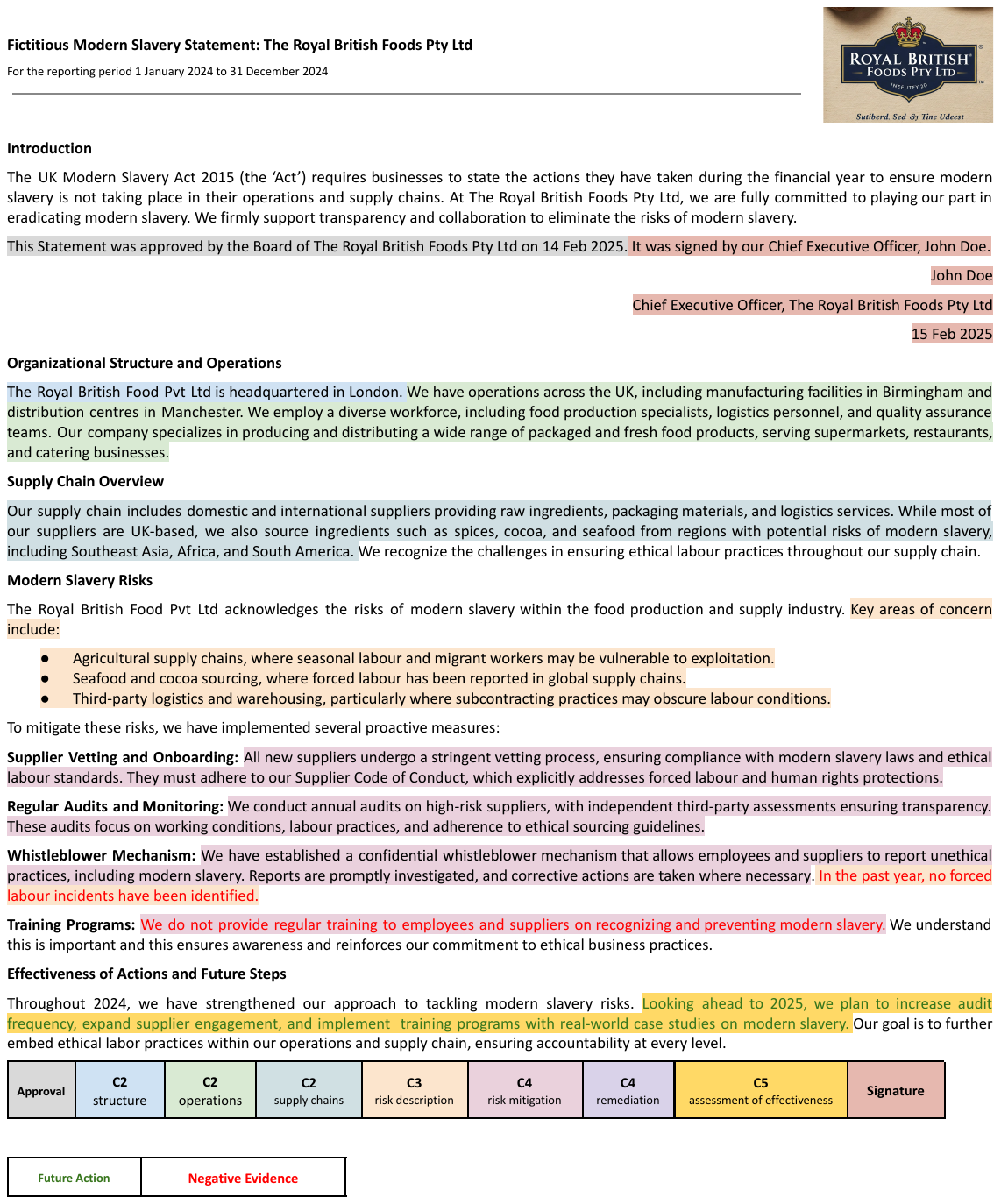}
     \caption{Example UK statement with nine criteria and evidence status annotations.}
     \label{fig:appendix_example_statement}
 \end{figure*}

\begin{figure*}[!htb]
    \centering
    \includegraphics[width=\linewidth]{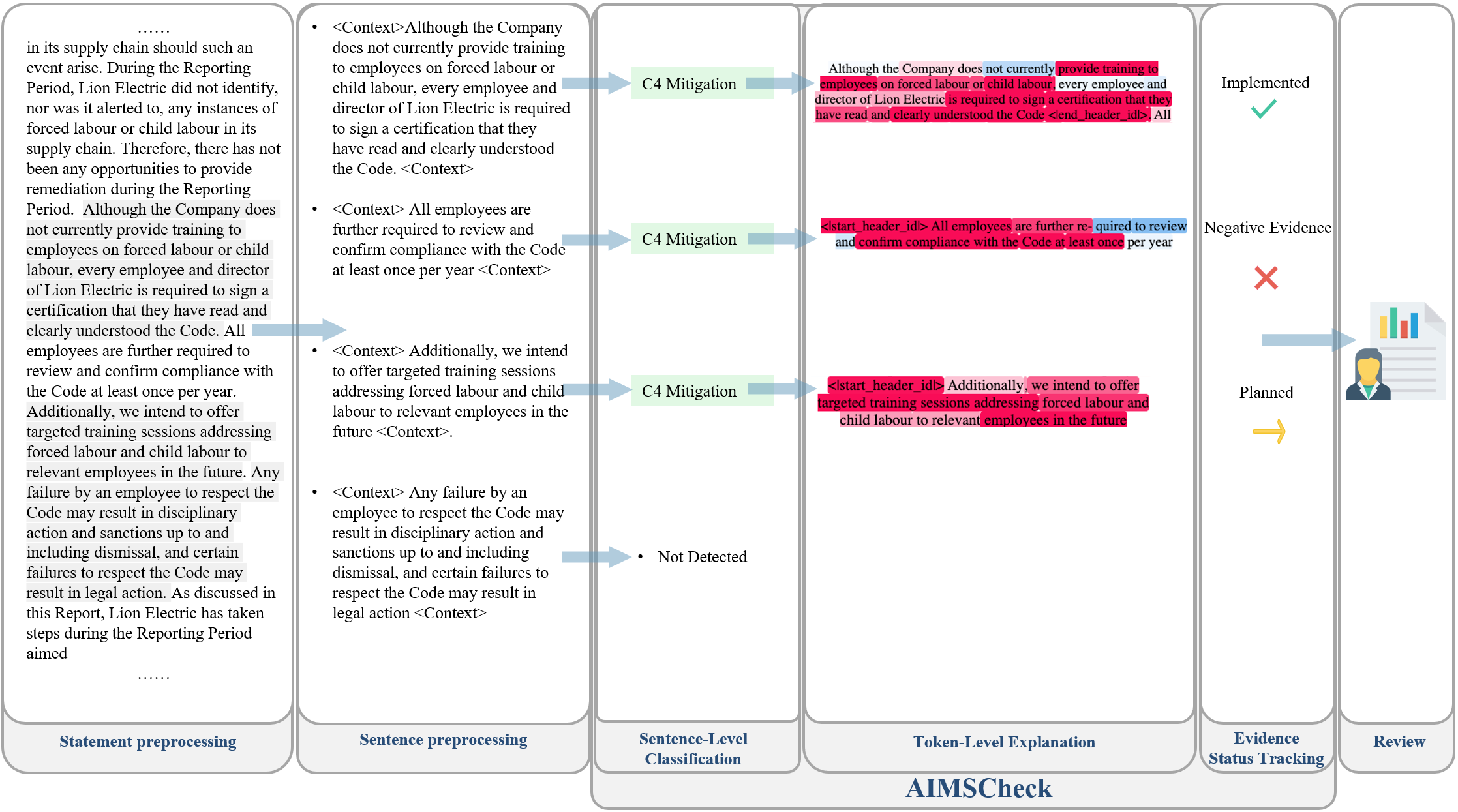}
    \caption{The processing workflow of a real statement example using the AIMSCheck framework. At statement preprocessing phase, the statement is split into sentences. The sentences are preprocessed to include surrounding context. The preprocessed sentences are passed an inputs to AIMSCheck. Within AIMSCheck, three levels of information are generated.  First, sentence-level classification where each sentence is classified per criteria. This is followed by token-level explanations for each classification. Third, evidence status is determined. Outputs from all three levels of AIMSCheck are then presented to a human reviewer who then determines the final compliance.}
    \label{fig:appendix_examples}
\end{figure*}

\newpage
\section{Example Statement with Annotations}
\label{sec:appendix_example_statement}
Figure \ref{fig:appendix_example_statement} shows an example UK statement with annotations for criteria and evidence status. The statements generally contain all the details about the company with focus on its risk accessment, supply chain and others as shown in the figure. We also include annotations for future promises (green). The explicit denial of action is also annotated (red).

\section{Real Statement Example of AIMSCheck Pipeline}
\label{appendix_realexample}

Figure~\ref{fig:appendix_examples} illustrates a real statement example processed through the AIMSCheck pipeline for analysing the mitigation criterion. The pipeline follows these key steps:

\paragraph{Step 1: Statement Preprocessing
}
The input consists of a statement parsed into individual sentences along with their neighboring contexts. This allows AIMSCheck to consider surrounding information when analysing each sentence.

\paragraph{Step 2: Sentence Preprocessing
}For the purpose of this example, a subset of four sentences is zoomed in for closer examination. To maintain brevity, omitted surrounding context is represented using the <context> symbol.

\paragraph{Step 3: Sentence-Level Classification
}AIMSCheck applies the fine-tuned Llama 3.2 model (100 context) on AIMS.au to classify sentences. In this example, the first three sentences meet the mitigation criterion, while the fourth does not.

\paragraph{Step 4: Token-Level Explanation
}Sentences identified as relevant to C4 mitigation undergo SHAP analysis. This step provides token-level explanations by highlighting the words that contribute most significantly to the model’s predictions.

\paragraph{Step 5: Evidence Status Tracking
}AIMSCheck then applies its Evidence Tracking methods to determine whether the identified mitigation evidence has already been implemented, is a negative evidence, or is planned for future execution.

\paragraph{Step 6:  Review}
Finally, all these results are provided to a user to manually review the evidence and explanations and make a final decision.

\section{Materials Release and License} All relevant materials, including datasets, model weights, notebooks, and prompts, are available on our project’s pages on 
\href{https://github.com/mila-ai4h/ai4h_aims-au}{GitHub}, 
\href{https://huggingface.co/datasets/mila-ai4h/AIMS.au}{Hugging Face}, and 
\href{https://figshare.com/articles/dataset/AIMS_au_Dataset/28489340}{Figshare}.
To promote broad accessibility while ensuring proper attribution, the materials are released under the Creative Commons Attribution 4.0 International (CC BY 4.0) license. This license permits unrestricted use, distribution, and reproduction in any medium, as long as the original authors and source are appropriately credited.

\onecolumn
\section{Mapping Legislations}
\label{appendix_mapping}

\begin{table}[h]
\label{mappingtable}
\centering
\resizebox{\textwidth}{!}{%
\begin{tabular}{p{6.5cm}p{5.5cm}p{6.5cm}}
\toprule
\textbf{The Australian Modern Slavery Act \cite{attorneygeneral2023}} & \textbf{Section 54(5) of The UK Modern Slavery Act \cite{homeoffice2024}} & \textbf{The Canadian Fighting Against Forced Labour and Child Labour in Supply Chains Act 
\cite{publicsafety2024}} \\
\midrule
Identify the reporting entity. 
& N/A 
& N/A \\
\hline
Describe the reporting entity’s structure, operations and supply chains. 
& Organisational structure, its business and its supply chains.
& Its structure, activities and supply chains. \\
\hline
Describe the risks of modern slavery practices in the operations and supply chains of the reporting entity and any entities it owns or controls.
& Assessing and managing risk.
& The parts of its business and supply chains that carry a risk of forced labour or child labour being used and the steps it has taken to assess and manage that risk. \\
\hline
Describe the actions taken by the reporting entity and any entities it owns or controls to assess and address these risks, including due diligence and remediation processes.
& Organisational policies, due diligence in relation to modern slavery (including approach to remediation), assessing and managing risk, training.
& Its policies and due diligence processes in relation to forced labour and child labour; any measures taken to remediate any forced labour or child labour; the training provided to employees on forced labour and child labour. \\
\hline
Describe how the reporting entity assesses the effectiveness of these actions.
& Monitoring and evaluation (understanding and demonstrating effectiveness).
& How the entity assesses its effectiveness in ensuring that forced labour and child labour are not being used in its business and supply chains. \\
\hline
Describe the process of consultation with any entities the reporting entity owns or controls (a joint statement must also describe consultation with the entity giving the statement).
& N/A 
& N/A \\
\hline
Provide any other relevant information.
& N/A 
& Any measures taken to remediate the loss of income to the most vulnerable families that results from any measure taken to eliminate the use of forced labour or child labour in its activities and supply chains. \\
\hline
Statement must be approved by the board.
& Approval from board of directors (or equivalent management body). 
& Approval by the organization’s governing body. \\
\hline
Signed by a responsible member of the organization.
& Signature from director or designated member. 
& Signature from members of governing body. \\
\bottomrule
\end{tabular}%
}
\caption{Comparative Mapping of Reporting Requirements of the Australian, UK, and Canadian Legislation}
\label{tab:legislation_mapping}
\end{table}
\twocolumn
\clearpage

\section{Experimental Settings}
\label{sec:experimental_setting}

\subsection{Model Fine Tuning}
\label{sec:appendix_models}
We fine-tuned two open models for the sentence relevance classification task, BERT \cite{DBLP:conf/naacl/DevlinCLT19} and Llama3.2 3B \cite{llama3herd2024}. The models were trained, starting from checkpoints available on HuggingFace repositories \cite{wolf2020huggingfacestransformersstateoftheartnatural}, using the annotated training data from the AIMS.au database \cite{bora2025}. For BERT, the full model weights were trained, while for Llama3.2 (3B) a LoRA approach was taken. Both models were trained for two different input setups, one where the sentence to be classified is provided without any surrounding text, and another where the sentence is surrounded by 100 words of context (balanced evenly before and after the sentence). All experiments were conducted on an A100L GPU with 80 GB of memory using PyTorch. BERT models were trained using the Adam optimizer and a learning rate of $3\times10^{-5}$. Llama models were trained using the Adam optimizer with a learning rate scheduler that includes 4000 warm up steps to a learning rate of $5\times10^{-4}$, followed by a $0.98$ decay every 400 steps. An effective batch size of 64 was used for all model training. The final model weights were selected to maximize the overal F1 score on a validation dataset. The BERT models were trained for 13h (no context) and 24h (100 words context), while the Llama models were trained for 35h (no context) and 41h (100 words context).

\subsection{GPT-4o and DeepSeek-R1}
We experiment with GPT-4o and DeepSeek-R1. We build upon the experiments conducted by \citet{bora2025}, leveraging the best-performing configurations identified in their study, GPT-4o with context of 100 words. We employ the same prompt structure as in their experiments and test the model across multiple datasets, including the AIMS.au, AIMS.ca, AIMS.uk. To further refine our approach, we conduct additional experiments by incorporating Chain of Thought (CoT) reasoning while maintaining the same context-based setup. These experiments were performed under both Zero-shot learning (no examples) and Few-shot learning (with a three to four examples per criteria). Finally, we replicate the same experiments using DeepSeek-R1, maintaining identical prompt structures to ensure comparability between models. Detailed examples of the prompts used in our experiments are provided in Appendix \ref{sec:appendix_prompts}. GPT-4o model experiments were conducted using the API from \cite{openaigpt4o}. Each criterion took between 8h to 15h to run, depending on the prompt, with a total of about \$1000 spent for the experiments. We setup 2.51bit quantized version of Deepseek-R1  640B using the implementation from \cite{unslothdeepseek}. The model was deployed using 4 H100 with 80 GB of memory using PyTorch and llama.cpp server. Each criteria took about 36 hours. 

\subsection{Evidence status tracking}
For future action detection, we developed a simple classifier using NLTK \cite{bird2006nltk} tense prediction with keywords such as "plan to" and "aim to." For negative evidence detection, we use a zero-shot BART-MNLI model \cite{lewis2019bart}. To improve sensitivity, we adjust the classification threshold from 0.5 to 0.35, as the model may lack confidence in predicting a sentence as a denial. The experiments were setup on a single GPU with 32GB memory.

\subsection{Prompts}
\label{sec:appendix_prompts}

In the below section, we show the sample prompts used in our experiments. Figure \ref{fig:appendix:prompt_design_and_ex:prompt_template_with_context} shows an example used in zero-shot GPT-4o and Deepseek-R1 experiments. Figure \ref{fig:appendix:prompt_design_and_ex:prompt_template_with_context0} shows CoT prompt and Figure \ref{fig:appendix:prompt_design_and_ex:prompt_template_with_context} shows few-shot prompt with examples.

\begin{figure*}
\begin{tcolorbox}[colback=gray!10!white, fontupper=\small, colframe=gray!80!black, title={Prompt template (C2, ``supply chains'', \emph{with-context})}]
You are an analyst that inspects modern slavery declarations made by Australian reporting entities. You are specialized in the analysis of statements made with respect to the Australian Modern Slavery Act of 2018, and not of any other legislation.
\newline\newline
You are currently looking for sentences in statements that describe the SUPPLY CHAINS of an entity, where supply chains refer to the sequences of processes involved in the procurement of products and services (including labour) that contribute to the reporting entity's own products and services. The description of a supply chain can be related, for example, to 1) the products that are provided by suppliers; 2) the services provided by suppliers, or 3) the location, category, contractual arrangement, or other attributes that describe the suppliers. Any sentence that contains these kinds of information is considered relevant. Descriptions that apply to indirect suppliers (i.e. suppliers-of-suppliers) are considered relevant. Descriptions of the supply chains of entities owned or controlled by the reporting entity making the statement are also considered relevant. However, descriptions of 'downstream' supply chains, i.e. of how customers and clients of the reporting entity use its products or services, are NOT considered relevant. Finally, sentences that describe how the reporting entity lacks information on some of its supply chain, or how some of its supply chains are still unmapped or unidentified, are also considered relevant.
\newline\newline
Given the above definitions of what constitutes a relevant sentence, you will need to determine if a target sentence is relevant or not inside a larger block of text. The target sentence will first be provided by itself so you can know which sentence we want to classify. It will then be provided again as part of the larger block of text it originally came from (extracted from a PDF file) so you can analyse it with more context. While some of the surrounding sentences may be relevant according to the earlier definitions, we are only interested in classifying the target sentence according to the relevance of its own content. You must avoid labeling sentences with only vague descriptions or corporate talk (and no actual information) as relevant.
\newline\newline
The answer you provide regarding whether the sentence is relevant or not can only be 'YES' or 'NO', and nothing else.
\newline\newline
The target sentence to classify is the following:
\newline
------------
\newline
\texttt{TARGET\_SENTENCE}
\newline
------------
\newline
The same target sentence inside its original block of text:
\newline
------------
\newline
\texttt{SENTENCE\_IN\_CONTEXT}
\newline
------------
\newline
Is the target sentence relevant? (YES/NO)
\end{tcolorbox}
\caption{Zero-shot prompt template used in model experiments under the \emph{with-context} setup.}
\label{fig:appendix:prompt_design_and_ex:prompt_template_with_context}
\end{figure*}

\begin{figure*}
\begin{tcolorbox}[colback=gray!10!white, fontupper=\small, colframe=gray!80!black, title={Chain of Thoughts Zero-Shot Prompt Template (C2, ``supply chains'', \emph{with-context})}]
You are an analyst specialising in the review of modern slavery declarations made by UK reporting entities under the UK Modern Slavery Act (the Act). Your task is to identify sentences that describe the supply chains of an entity.
\newline\newline
Key Rules:
Supply chains refer to the sequences of processes involved in the procurement of products and services (including labour) that contribute to the reporting entity’s own products and services. 
\newline\newline
**Relevant sentences** may include descriptions such as: 

\begin{itemize}
    \item The products that are provided by suppliers.
    \item The services provided by suppliers.
    \item The location, category, contractual arrangement, or other attributes that describe the suppliers.
    \item Descriptions related to indirect suppliers (i.e., suppliers-of-suppliers).
    \item Descriptions of the supply chains of entities owned or controlled by the reporting entity making the statement.
    \item Information about how the reporting entity lacks information on some of its supply chain, or how some of its supply chains are still unmapped or unidentified.
\end{itemize}
**Irrelevant sentences** may include vague statements about suppliers without specific descriptions, or sentences describing downstream supply chains (i.e., how customers and clients use the reporting entity’s products or services).
Task:
You will be given a target sentence and its surrounding context from a modern slavery statement. Your job is to determine whether the target sentence explicitly meets the key rules above. Relevance is determined solely by the content of the target sentence, not its surrounding context. Your answer must be YES or NO. 
Now classify the following target sentence:

The target sentence to classify is the following:
\newline
------------
\newline
\texttt{TARGET\_SENTENCE}
\newline
------------
\newline
The same target sentence inside its original block of text:
\newline
------------
\newline
\texttt{SENTENCE\_IN\_CONTEXT}
\newline
------------
\newline

**Question**:
Is the target sentence relevant? (YES/NO)

**Answer**: Lets think step-by-step. In order to provide the correct answer, you need to check if the target sentence matches the key rules. Provide the reasoning and the final answer (YES or NO).

Reasoning:

Final Answer: YES/NO"""

\end{tcolorbox}
\caption{Chain-of-Thoughts zero-shot prompt template used for zero-shot model experiments under the \emph{with-context} setup.}
\label{fig:appendix:prompt_design_and_ex:prompt_template_with_context0}
\end{figure*}

\begin{figure*}
\begin{tcolorbox}[colback=gray!10!white, fontupper=\tiny, colframe=gray!80!black, title={Chain of Thoughts Few-Shot Prompt Template (C2, ``supply chains'', \emph{with-context})}]
You are an analyst specialising in the review of modern slavery declarations made by UK reporting entities under the UK Modern Slavery Act (the Act). Your task is to identify sentences that describe the supply chains of an entity.
\newline\newline
Key Rules:
Supply chains refer to the sequences of processes involved in the procurement of products and services (including labour) that contribute to the reporting entity’s own products and services. 
\newline\newline
**Relevant sentences** may include descriptions such as: 

\begin{itemize}
    \item The products that are provided by suppliers.
    \item The services provided by suppliers.
    \item The location, category, contractual arrangement, or other attributes that describe the suppliers.
    \item Descriptions related to indirect suppliers (i.e., suppliers-of-suppliers).
    \item Descriptions of the supply chains of entities owned or controlled by the reporting entity making the statement.
    \item Information about how the reporting entity lacks information on some of its supply chain, or how some of its supply chains are still unmapped or unidentified.
\end{itemize}
**Irrelevant sentences** may include vague statements about suppliers without specific descriptions, or sentences describing downstream supply chains (i.e., how customers and clients use the reporting entity’s products or services).
\newline\newline
Task:
You will be given a target sentence and its surrounding context from a modern slavery statement. Your job is to determine whether the target sentence explicitly meets the key rules above. Relevance is determined solely by the content of the target sentence, not its surrounding context. Your answer must be YES or NO. 

Examples with reasoning: 

\textbf{Example 1:}
\begin{itemize}
    \item Target Sentence: "Our supply chain includes providers of remediation products and other project-focused materials purchased and distributed through Duratec warehouses or delivered directly to project sites through third parties. Products are purchased domestically and imported through third-party logistics providers. Our suppliers are located principally in the UK and at least 12 foreign countries."
    \item Question: Is the target sentence relevant? (YES/NO)
    \item Reasoning: This sentence provides specific information about the types of products in the supply chain, the distribution methods, the geographical locations of suppliers, and the involvement of third-party logistics providers. These details directly describe the entity's supply chain as defined by the key rules.
    \item Final Answer: YES
\end{itemize}

\textbf{Example 2:}
\begin{itemize}
    \item Target Sentence: "We procure goods and services from trusted suppliers across the world."
    \item Question: Is the target sentence relevant? (YES/NO)
    \item Reasoning: While this sentence mentions procuring goods and services from suppliers globally, it is too vague and lacks specific details about the suppliers, their products or services, locations, or other attributes that describe the supply chain.
    \item Final Answer: NO
\end{itemize}

\textbf{Example 3:}
\begin{enumerate}
    \item Target Sentence: "We continue due diligence across all our direct and indirect suppliers.”
    \item Question: Is the target sentence relevant? (YES/NO)
    \item Reasoning: This sentence mentions due diligence efforts but does not provide specific descriptions of suppliers, their products or services, locations, or other attributes. It lacks the detailed information required to describe the supply chain.
    \item Final Answer: NO
\end{enumerate}

Now classify the following target sentence:

The target sentence to classify is the following:
\newline
------------
\newline
\texttt{TARGET\_SENTENCE}
\newline
------------
\newline
The same target sentence inside its original block of text:
\newline
------------
\newline
\texttt{SENTENCE\_IN\_CONTEXT}
\newline
------------
\newline

**Question**:
Is the target sentence relevant? (YES/NO)

**Answer**: Lets think step-by-step. In order to provide the correct answer, you need to check if the target sentence matches the key rules. Provide the reasoning and the final answer (YES or NO).

Reasoning:

Final Answer: YES/NO"""

\end{tcolorbox}
\caption{Chain-of-Thoughts few-shot prompt template used for zero-shot model experiments under the \emph{with-context} setup.}

\label{fig:appendix:prompt_design_and_ex:prompt_template_with_context}
\end{figure*}

\subsection{Other datasets} We conducted preliminary experiments on the WikiRate \cite{WikiRate2023} dataset and found that mapping AIMS.au questions to WikiRate criteria is feasible. However, extracting sentence-level annotations from WikiRate data proved more difficult. Exploring this publicly available annotated dataset further, presents promising directions for future work.

\clearpage
\subsection{Mapping other jurisdictions}
\label{sec:other_jurisdictions}

Table \ref{tab:other_legislations} demonstrates the feasibility of mapping between the criteria used in AIMSCheck across mHRRD legislative frameworks. Although compliance criteria frequently address similar procedural elements, such as risk identification, their specific focus often diverges, for instance, targeting modern slavery in some cases versus broader human rights concerns in others. AIMSCheck partially accounts for these variations, as evidenced by the ability of models trained on the Australian Act’s modern slavery-focused language to generalize effectively to related themes such as forced labor and child labor in the Canadian Act. These initial findings point to promising avenues for further investigation in future work.

\begin{table*}[h]
\centering
\renewcommand{\arraystretch}{1.2}
\resizebox{\textwidth}{!}{%
\begin{tabular}{@{}>{\raggedright\arraybackslash}p{4.5cm}
                >{\raggedright\arraybackslash}p{5.8cm}
                >{\raggedright\arraybackslash}p{5.8cm}
                >{\raggedright\arraybackslash}p{5.8cm}@{}}
\toprule
\textbf{AIMSCheck Criteria} & \textbf{France's Duty of Vigilance Law
  \cite{france2017official}} & \textbf{Act on Corporate Due Diligence in Supply Chains \cite{germany2021}} & \textbf{Norwegian The Transparency Act \cite{forbrukertilsynet2022}} \\
\midrule
Approval \& Signature & Common practice (but not legally required) & Common practice (but not legally required) & Signed per Section 3-5 of the Accounting Act. \\
\hline
C2: Structure, Operations, Supply Chains & Implicitly expected (but not legally required)& Implicitly expected (but not legally required) & A general description of the enterprise’s structure, area of operations, guidelines and procedures for handling actual and potential adverse impacts on fundamental human rights and decent working conditions. \& operations \\
\hline
C3: Risk Description & A mapping that identifies, analyses and ranks risks. & Whether and which human rights and environment-related risks the enterprise has identified. & Information regarding actual adverse impacts and significant risks of adverse impacts that the enterprise has identified through its due diligence.  \\
\hline
C4: Risk Mitigation \& Remediation & Procedures to regularly assess, in accordance with the risk mapping, the situation of subsidiaries, subcontractors or suppliers with whom the company maintains an established commercial relationship. Appropriate action to mitigate risks or prevent serious violations. An alert mechanism that collects reporting of existing or actual risks, developed in working partnership with the trade union organizations representatives of the company concerned. & What the enterprise has done to fulfil its due diligence obligations. & Information regarding measures the enterprise has implemented or plans to implement to cease actual adverse impacts or mitigate significant risks of adverse impacts, and the results or expected results of these measures.  \\
\hline
C5: Effectiveness & A monitoring scheme to follow up on the measures implemented and assess their efficiency & How the enterprise assesses the impact and effectiveness of the measures. What conclusions it draws for future measures. & Results of implemented measures \\
\bottomrule
\end{tabular}%
}
\caption{Mapping AIMSCheck Criteria to Other Jurisdictions}
\label{tab:other_legislations}
\end{table*}

\subsection{Results}
We show the per category results for GPT-4o experiments in Tables~\ref{tab:gpt4o}, \ref{tab:cot_zeroshot} and \ref{tab:cot_fewshot}.
\begin{table*}[!htb]
    \centering
    \begin{adjustbox}{max width=\textwidth}
    \begin{tabular}{lccccccccc}
  \toprule
        & \multicolumn{9}{c}{\textbf{GPT-4.o Results}} \\
        \cmidrule(lr){2-10}
        \textbf{Category} & \textbf{Approval} & \textbf{Signature} & \textbf{Structure} & \textbf{Operations} & \textbf{Supply Chains} & \textbf{Risks} & \textbf{Mitigation} & \textbf{Remediation} & \textbf{Assessment} \\
        \midrule
        AIMS.au & 0.754 & 0.477 & 0.723 & 0.613 & 0.557 & 0.563 & 0.698 & 0.562 & 0.459 \\
        AIMS.uk & 0.489 & 0.515 & 0.579 & 0.608 & 0.485 & 0.472 & 0.714 & 0.729 & 0.289 \\
        AIMS.ca  & 0.723 & 0.682 & 0.677 & 0.614 & 0.444 & 0.406 & 0.664 & 0.567 & 0.459 \\
        Wikirate & 0.623 & 0.699 & 0.452 & 0.241 & 0.104 & 0.258 & 0.398 & 0.415 & 0.198 \\
        \bottomrule
    \end{tabular}
    \end{adjustbox}
    \caption{F1 Scores for GPT-4.o for each criterion shown for all three benchmark datasets.}
    \label{tab:gpt4o}
\end{table*}

\begin{table*}[!htb]
    \centering
    \begin{adjustbox}{max width=\textwidth}
    \begin{tabular}{lccccccccc}
    \toprule
        & \multicolumn{9}{c}{\textbf{GPT-4.o CoT Zero-Shot}} \\
        \cmidrule(lr){2-10}
        \textbf{Category} & \textbf{Approval} & \textbf{Signature} & \textbf{Structure} & \textbf{Operations} & \textbf{Supply Chains} & \textbf{Risks} & \textbf{Mitigation} & \textbf{Remediation} & \textbf{Assessment} \\
        \midrule
        AIMS.au & 0.855 & 0.409 & 0.619 & 0.529 & 0.420 & 0.422 & 0.709 & 0.552 & 0.518 \\
        AIMS.uk  & 0.634 & 0.477 & 0.573 & 0.510 & 0.332 & 0.283 & 0.722 & 0.672 & 0.296 \\
        AIMS.ca & 0.839 & 0.667 & 0.668 & 0.512 & 0.348 & 0.374 & 0.680 & 0.529 & 0.419 \\
        Wikirate & 0.753 & 0.571 & 0.397 & 0.179 & 0.058 & 0.141 & 0.370 & 0.393 & 0.199 \\
        \bottomrule
    \end{tabular}
    \end{adjustbox}
    \caption{F1 Scores for GPT-4.o CoT for each criterion shown for all three benchmark datasets.}
    \label{tab:cot_zeroshot}
\end{table*}

\begin{table*}[!htb]
    \centering
    \begin{adjustbox}{max width=\textwidth}
    \begin{tabular}{lccccccccc}
    \toprule
        & \multicolumn{9}{c}{\textbf{GPT-4.o CoT Few-Shot}} \\
        \cmidrule(lr){2-10}
        \textbf{Category} & \textbf{Approval} & \textbf{Signature} & \textbf{Structure} & \textbf{Operations} & \textbf{Supply Chains} & \textbf{Risks} & \textbf{Mitigation} & \textbf{Remediation} & \textbf{Assessment} \\
        \midrule
        AIMS.au & 0.843 & 0.636 & 0.658 & 0.651 & 0.556 & 0.450 & 0.664 & 0.601 & 0.492 \\
        AIMS.uk  & 0.650 & 0.724 & 0.564 & 0.645 & 0.461 & 0.346 & 0.695 & 0.672 & 0.398 \\
        AIMS.ca & 0.912 & 0.732 & 0.689 & 0.634 & 0.461 & 0.434 & 0.664 & 0.561 & 0.443 \\
        Wikirate & 0.750 & 0.659 & 0.433 & 0.249 & 0.096 & 0.165 & 0.374 & 0.444 & 0.174 \\
        \bottomrule
    \end{tabular}
    \end{adjustbox}
    \caption{F1 Scores for GPT-4.o CoT few-shots for each criterion shown for all three benchmark datasets.}
    \label{tab:cot_fewshot}
\end{table*}

\section{Calibration Curves}
\label{sec:appendix_calibration}
\begin{figure*}
    \centering
    \includegraphics[width=\linewidth]{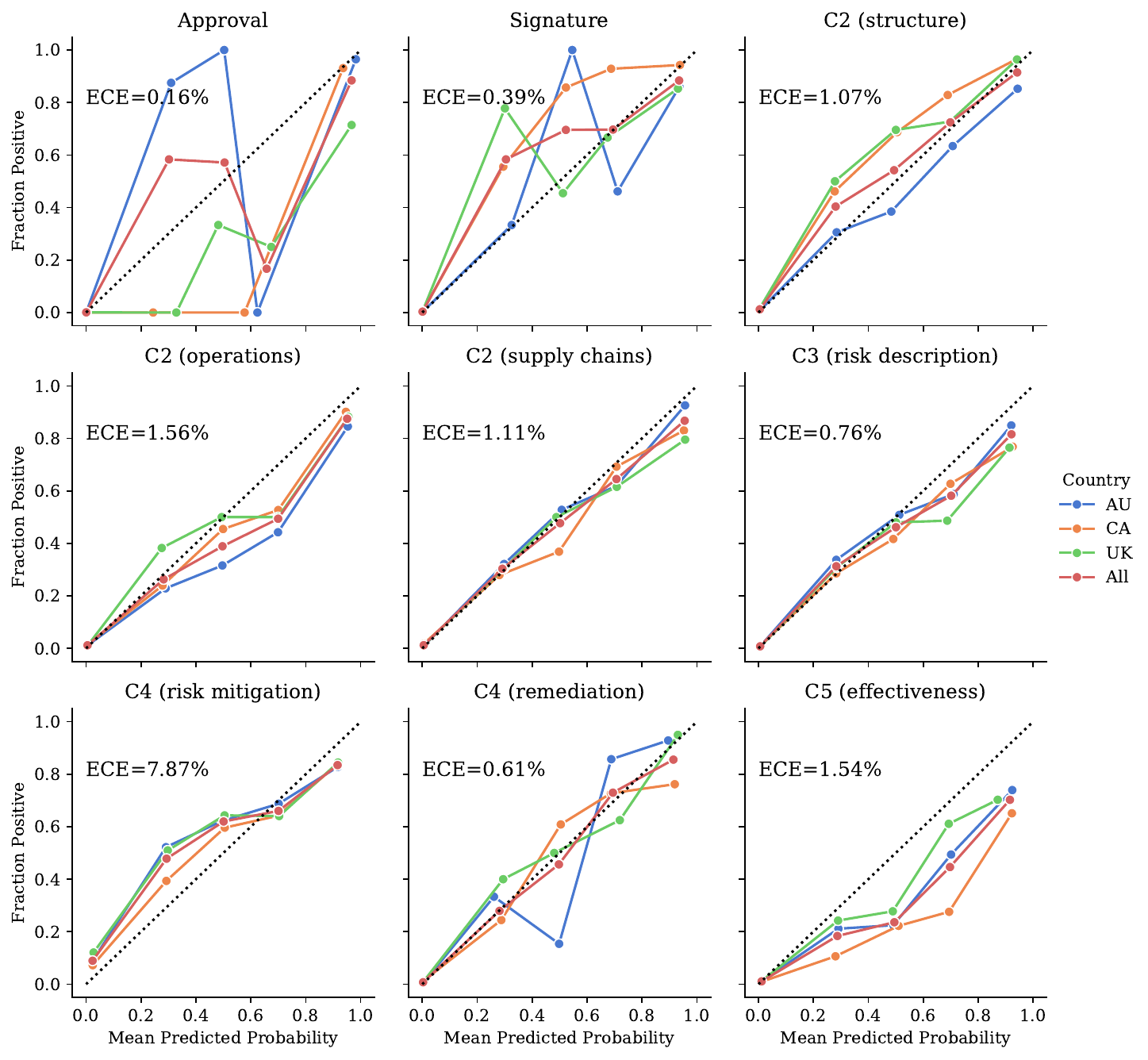}
    \caption{Calibration curves for the Llama3.2 3B model with 100 context words. The dashed black lines correspond to a perfectly calibrated model where the predicted probabilities match the fraction of samples that are positive. The Expected Calibration Error (ECE) metric for all countries is also shown alongside the curves. The calibration curves were computed using five uniformly spaced bins, while the ECE metrics used ten bins.}
    \label{fig:calibration_curves}
\end{figure*}

Figure \ref{fig:calibration_curves} shows calibration curves for the Llama3.2 3B model with 100 context words. Overall, the fine-tuned model is well calibrated, meaning that prediction probabilities output by the model are closely indicative of how often the model will be correct in its prediction. We note that calibration curves and Expected Calibration Error (ECE) are affected when data is highly imbalanced, as is the case for much of our data. As shown in Figure \ref{fig:relevance_ratios}, the most imbalanced class is "Approval", and the least imbalanced class is "C4 (risk mitigation)". Respectively, these classes have the lowest and highest ECE values.

\section{SHAP Visualizations}
\label{sec:shap}
\begin{figure*}[htbp]
    \centering
        \begin{subfigure}[b]{\linewidth}
        \centering
        \includegraphics[width=\linewidth]{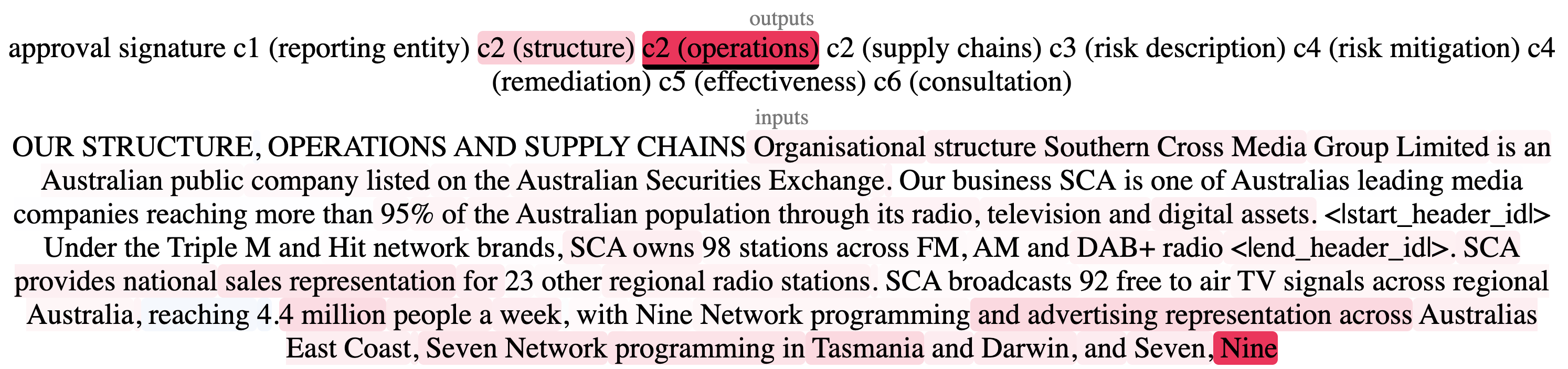}
        \caption{This example show the SHAP plot for label C2 operations (underlined). C2 operations has a high prediction value (intense red). There are no blue tokens in the text that would negatively influence the prediction.}
        \label{fig:shap2}
    \end{subfigure}%
    \vspace{2cm}
    \begin{subfigure}[b]{\linewidth}
        \centering
        \includegraphics[width=\linewidth]{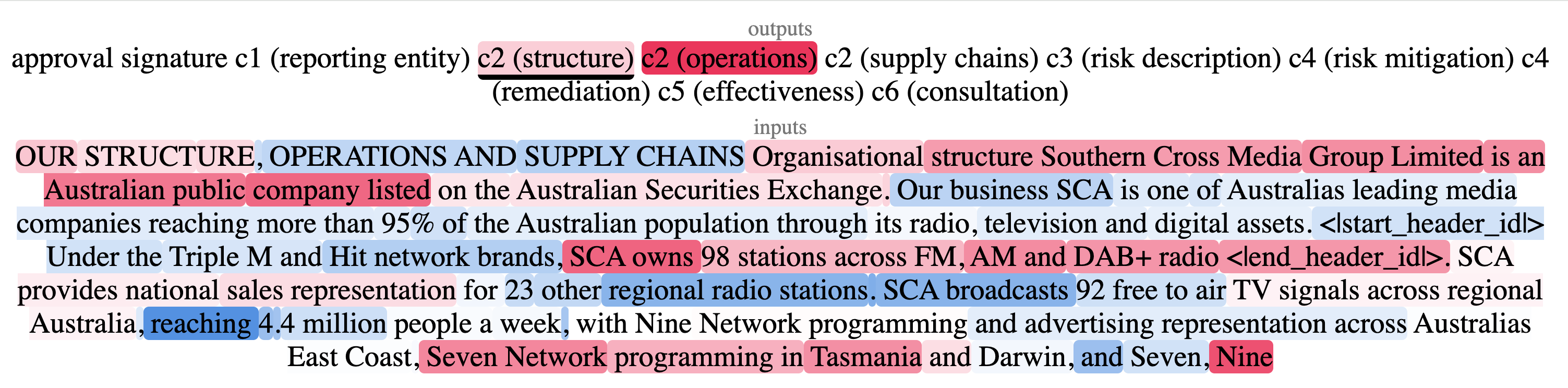}
        \caption{This example show the SHAP plot for label C2 structure (underlined). SHAP plot illustrates the mixed influence of both negative and positive tokens on the model's output. The interpretation of the model is guided by the opposing contributions from these conflicting factors.}
        \label{fig:shap1}
    \end{subfigure}%
    \vspace{2cm}

    \begin{subfigure}[b]{\linewidth}
        \centering
        \includegraphics[width=\linewidth]{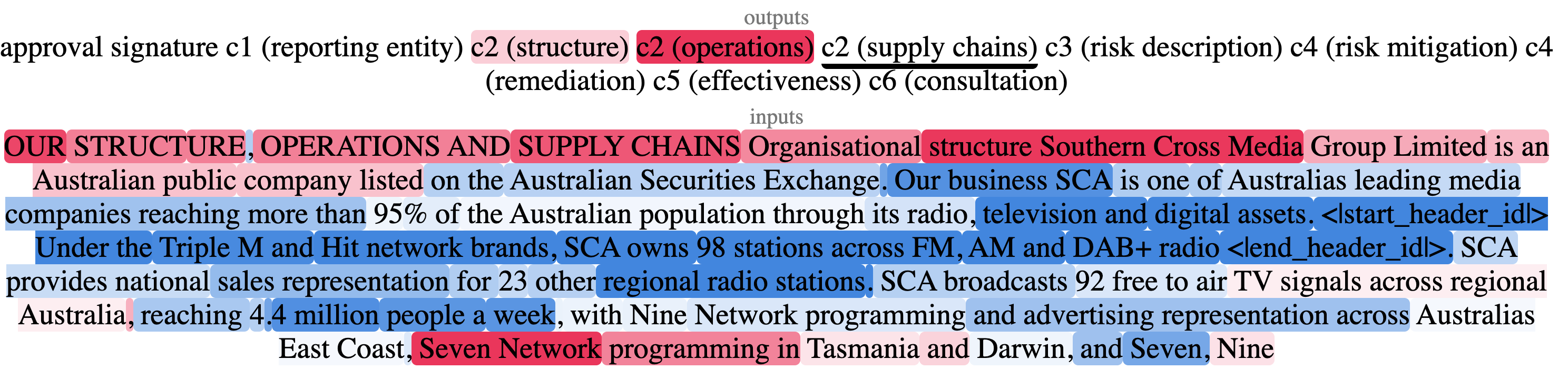}
        \caption{This example show the SHAP plot for label c2 supply chains (underlined). Many blue tokens indicate negative influence on the model output. This resulted in a negative prediction for C2 supply chains.}
        \label{fig:shap3}
    \end{subfigure}
    \vspace{0.5cm}
    \caption{Visualization of SHAP values for the target sentence (encompassing the special header tags) with surrounding 100 context words. All the criteria labels are shown on top of each figure. In the figure, we show SHAP plots for three different labels on the same target example. The plots differ by classification label-the selected label is underlined in each figure. Red regions correspond to parts
    of the text that increase the output of the model when they are included, while blue regions decrease the output of
    the model when they are included. The intensity of the color indicates the strength of the influence. The model predicted labels C2 operations and C2 structure for this example (labels in red).
   \label{fig:shap_all}}
\end{figure*}

We provide additional examples of SHAP text plots in Figure \ref{fig:shap_all}.

\clearpage
\section{Error Analysis of Llama3.2 and GPT-CoT Few-Shot Models}
\label{error_analysis}
This appendix provides a comprehensive breakdown of the error patterns observed in our evaluation of Llama3.2 and GPT-CoT Few-Shot on the AIMS.uk and AIMS.ca. Our findings indicate no substantial differences between the datasets in terms of model error distributions. Instead, challenges stem from inherent model tendencies, formatting issues, and semantic overlaps in legal language.

\subsection{False Positives
}

GPT-CoT frequently over-predicts, leading to a higher false positive rate. Despite explicit prompts, it occasionally selects vague or misleading sentences. Example:
\begin{itemize}
    \item “In 2020, the company’s Human Rights Policy was approved by the Board of Directors.”
The presence of "approved" misleads the model into incorrectly classifying the sentence under the approval criterion, even though it does not refer to a modern slavery statement.
\end{itemize} 
What is more, the model often relies on contextual relevance rather than assessing sentences independently, despite explicit instructions to base predictions solely on the target sentence. This tendency leads to frequent false positives.

\paragraph{Confusion Between Closely Related Criteria} One of the most prevalent error types for both models, arises from the models' difficulty in distinguishing between overlapping legal criteria. For example, elements of Criterion 2 (Structure, Operations, and Supply Chains) are frequently misclassified between them.
\begin{itemize}
    \item Example:“Jadestone is an independent oil and gas company focused on mid-life production and near-term development assets in the Asia Pacific region, operating principally in Australia, Indonesia and Malaysia.”. The sentence includes relevant information about operations, but the Llama 3.2 model classifies it as structure.
\end{itemize}

\paragraph{Annotation Inconsistencies} A minor fraction (1–2\%) of false positives stems from annotation inconsistencies, where the human annotator occasionally mislabel certain entries. These errors contribute to noise in model predictions.

\subsection{False Negatives}

\paragraph{Formatting Issues} The models struggle with  poorly extracted list and segmented sentences.
\begin{itemize}
    \item List Extraction Errors: It may omit entities when lists span multiple sentences (e.g., sentence 1: “Our subsidiaries include” sentence 2: “Company A”, sentence 3: “Company B”, sentence 4: “and Company C.”).  
\end{itemize}
\begin{itemize}
    \item Broken Sentences: Splitting key phrases across lines can hinder recognition. This is a common error for predicting the signature where the name, title, and signature indicator are extracted in separate sentences. 
\end{itemize}

\paragraph{Overlapping Legal Criteria} The models struggle to differentiate between closely related criteria, often predicting only one when multiple apply. 

\begin{itemize}

    \item Example: “With offices in 8 countries and approximately 2,300 seagoing and shore-based employees, we provide a comprehensive set of marine services to the world's leading energy companies.” This sentence includes evidence for both structure (number of employees) and operations (services), but the Llama 3.2 model assigns it only to operations. 
\end{itemize}
This type of confusions are expected, as legal requirements often group related aspects under a single question while the annotated data looked at each of them separate (see Figure \ref{fig:data_mapping}). Companies frequently address these together, using mixed terminology, making it difficult for the model to distinguish them.

\paragraph{Omission of Key Relevant Sentances} Llama3.2 demonstrates a tendency to be conservative missing key relevant information such as headquarters location or employee count when assessing sentences for structure criterion or risk assessments for mitigation criterion. However, despite overpredicting, the GPT model also seems to miss key relevant information just as Lllama 3.2. 
\begin{itemize}
    \item Example:“This statement was approved by the Board of Directors on January 1, 2024.”  The GPT model assigns an incorrect negative label despite clear approval language.
\end{itemize}

What is more, both models miss many relevant sentences that contains negative evidence. 
\begin{itemize}
    \item Example:“During the calendar year of 2023, no concerns of child labour were identified and therefore no remedial measures were undertaken”. This sentence is relevant both for the risk description criterion and for the remediation criterion.
\end{itemize}

\subsection{Future Directions}

The observed error patterns highlight the importance of refining the model prompts, improving the consistency of the data annotation, and developing advanced extraction strategies to enhance precision. Future work should focus on:

\begin{enumerate}
    \item Better sentence segmentation techniques and improve list extraction to reduce errors caused by formatting inconsistencies.
    \item Enhanced prompt engineering to specifically address false positives.
    \item Additional dataset augmentation to reduce confusion in overlapping criteria and the false negatives in fine-tuning models.
\end{enumerate}

\section{Compliance Trend in the AIMS.uk Dataset}
\label{sec:complianceTrend}

In this appendix, we present a compliance trend analysis based on the prediction outcomes of the Llama3 context-100 model on the 50 statements of AIMS.uk. For each criterion where a statement was classified as positive, we analyse the compliance proportion concerning \textit{corporate turnover}, \textit{sector}, and the \textit{statement's publication year}.

We hope this analysis serves as an example of the insights our models can generate, enabling more thorough compliance assessments, particularly when applied at scale. As of February 2025, the UK Modern Slavery Register \cite{ukgovmodernslaveryregistry} contains 57,594 statements, while the Canadian Registry \cite{canadianmodernslaveryact} includes 6,366. Future research could focus on expanding this analysis to cover all available statements.

\subsection{Turnover}
\begin{figure*}[htb]
    \centering
    \includegraphics[width=0.8\textwidth, height=0.4\textwidth,trim={0 0px 0 30px},clip]{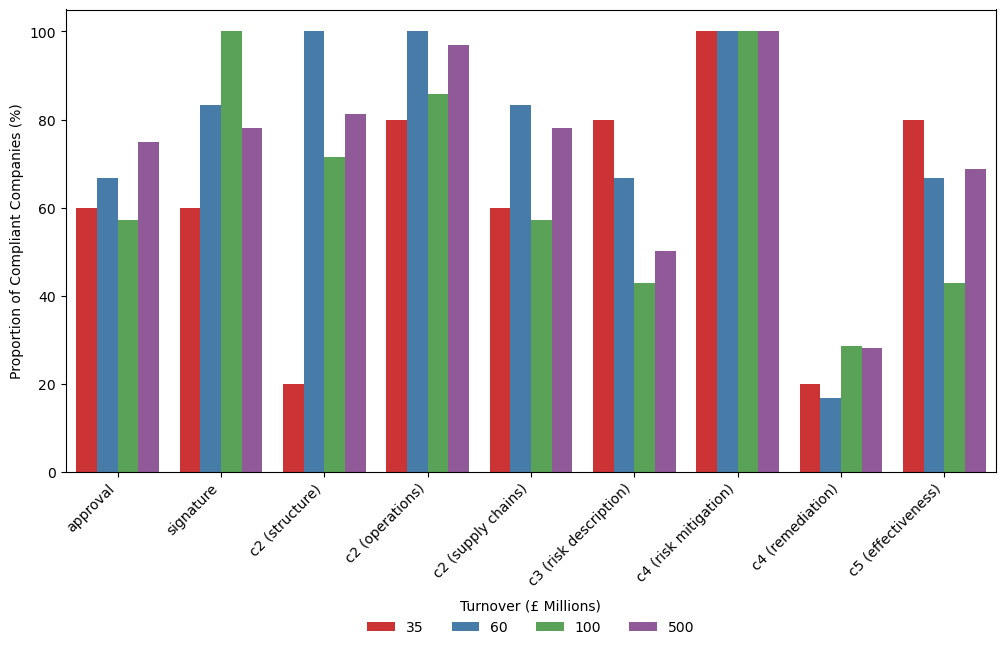}
    \caption{Compliance Proportion by Turnover (£ Millions). for Each Criterion}
    \label{fig:turnover-apd}
\end{figure*}

We first examined the turnover distribution across the entire AIMS.uk dataset and we calculated the proportion of companies compliant with each criterion within different turnover categories. Figure~\ref{fig:turnover-apd} illustrates the compliance proportion across turnover categories for each criterion. 

We opted to analyze companies with an annual turnover of less than £36 million as a distinct category. The UK MSA mandates reporting for commercial organizations operating in the UK with a turnover of £36 million or more, while those below this threshold are not legally required to report but may do so voluntarily \cite{ModernSlaveryAct2015}. This distinction provides a valuable comparison point for our analysis.

\paragraph{Key Observations}
\begin{itemize}
    \item Smaller companies, publishing voluntary statements, are outperforming larger ones in detailing their risk description and evaluating the effectiveness of their actions. This trend may suggest that larger companies are either hesitant to disclose their findings or face challenges in mapping their risks due to their size.
\end{itemize}

\begin{itemize}
    \item It's noteworthy that remediation is the least addressed category among companies. However, it is encouraging to observe that larger companies are articulating their actions to remediate potential incidents.
\end{itemize}

\subsection{Sector}

We categorize industries into four broad sectors: 
Industry \& Infrastructure, Commerce \& Services, Public \& Healthcare, and Other. 

\begin{enumerate}
    \item The Industry \& Infrastructure sector includes industries such as automotive, construction, durable consumer goods, mining, utilities, waste management, transportation, defense, and security services.
    \item The Commerce \& Services sector covers consumer services, including hospitality, tourism, fashion, cosmetics, food and agriculture, finance, professional services, IT, and media.
    \item The Public \& Healthcare sector consists of public sector organizations, non-profits, education, healthcare, and forestry-related industries.
    \item The Other category includes the statements that companies that had "other" in their metdata.
\end{enumerate}

\begin{figure*}[htb]
    \centering
     \includegraphics[width=0.8\textwidth, height=0.4\textwidth,trim={0 0px 0 30px},clip]{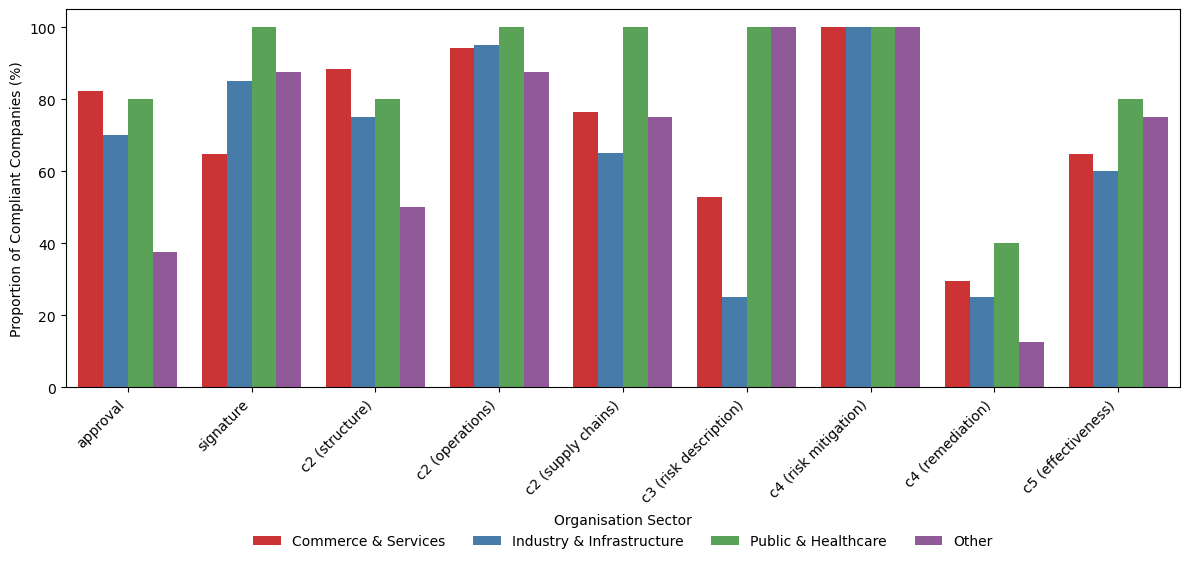}
    \caption{Compliance Proportion by Sector for Each Criterion}
    \label{fig:sector-apd}
 \end{figure*}

Figure~\ref{fig:sector-apd} shows the compliance proportion by sector for each criterion.

\paragraph{Key Observations}

\begin{itemize}
    \item Companies within the Public and Healthcare sectors demonstrate the highest levels of compliance with all mandatory disclosure criteria.
    \item Conversely, firms operating in the Industry and Infrastructure sectors appear to be less forthcoming in their risk disclosures. This relative opacity may indicate that these companies experience lower external pressures to disclose risks compared to more public-facing industries, such as commerce and healthcare.
\end{itemize}

\subsection{Statement Publication Year}
\begin{figure*}[!htb]
    \centering
    \includegraphics[width=0.8\textwidth, height=0.4\textwidth,trim={0 0px 0 30px},clip]{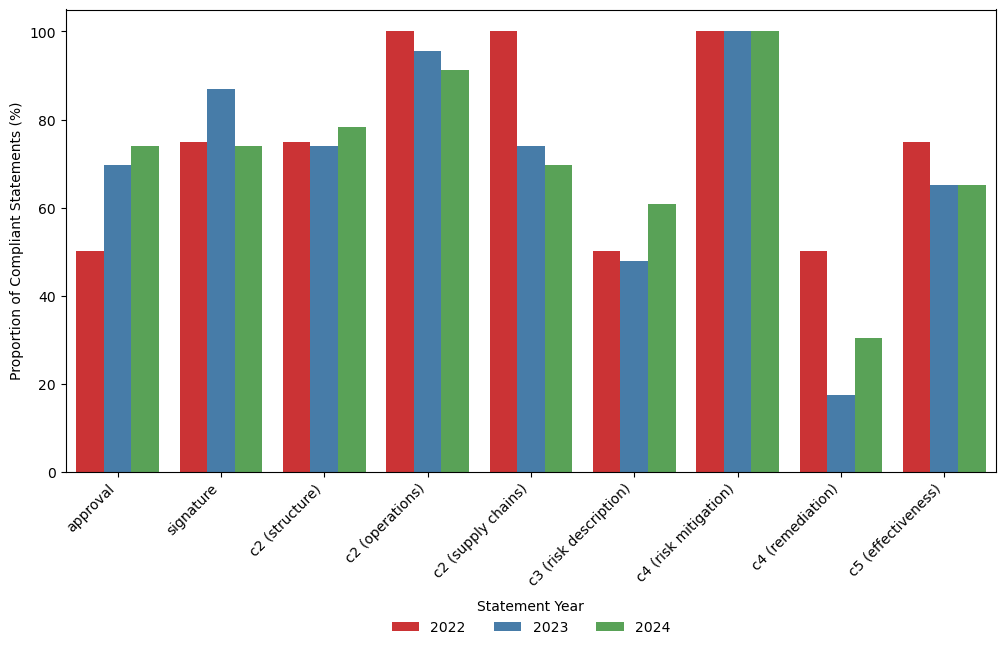}
    \caption{Compliance Proportion by Publication Year for Each Criterion}
    \label{fig:year-apd}
\end{figure*}

Similarly, we computed the compliance proportion based on the statement publication year and then illustrate in Figure~\ref{fig:year-apd} the compliance trend over different publication years for each criterion. 

\paragraph{Key Observations}
While it could be expected that the companies are getting better at reporting over years, we see a trend of more companies failing to disclose their operations and supply chain from 2022 to 2024. 


\end{document}